\documentclass[a4paper]{article}
\usepackage[utf8]{inputenc}
\usepackage[T1]{fontenc}
\usepackage{graphicx}
\usepackage[textwidth=3.7cm, textsize=scriptsize]{todonotes}
\usepackage{caption}
\usepackage{subcaption}
\usepackage{booktabs}
\usepackage{multirow, array}
\usepackage{color, soul}
\usepackage{hyperref}
\usepackage{xspace}
\usepackage{paralist}
\usepackage{authblk}
\usepackage{natbib}
\usepackage[acronym]{glossaries}

\newacronym{osm}{OSM}{OpenStreetMap}
\newacronym{dem}{DEM}{Digital Elevation Model}
\newacronym{dtm}{DTM}{Digital Terrain Model}
\newacronym{gtfs}{GTFS}{General Transit Feed Specification}
\newacronym{matsim}{MATSim}{Multi-Agent Transport Simulation}
\newacronym{vista}{VISTA}{Victorian Integrated Survey for Travel and Activity}
\newacronym{pt}{PT}{Public Transportation}
\newacronym{iv}{IV}{Infrastructure Victoria}
\newacronym{fn}{FN}{Full Network}
\newacronym{sn}{SON}{Standard Output Network}
\newacronym{cn}{CN}{Concise Network}
\newacronym{mn}{MGN}{MATSim-Generated Network}
\newacronym{ivn}{IVN}{Infrastructure Victoria Network}

\newcommand{\ms}{\acrshort{matsim}\xspace}
\newcommand{\osm}{\acrshort{osm}\xspace}

\newcommand{\sn}[0]{\gls{sn}\xspace}

\newcommand{\mn}[0]{\gls{mn}\xspace}
\newcommand{\ivn}[0]{\gls{ivn}\xspace}
\newcommand{\n}[0]{$\mathcal{N}$\xspace}
\newcommand{\comment}[1]{}

\providecommand{\keywords}[1]
{
  \small	
  \textbf{\textit{Keywords---}} #1
}

\title{Building the road network for city-scale active transport simulation models}
\author[1]{Afshin Jafari}
\author[1]{Alan Both}
\author[2,3]{Dhirendra Singh}
\author[1]{Lucy Gunn}
\author[1]{Billie Giles-Corti}

\affil[1]{School of Global, Urban and Social Studies, RMIT University}
\affil[2]{School of Computing Technologies, RMIT University}
\affil[3]{Data61, CSIRO}

\begin{document}

\maketitle

\begin{abstract}

   In this paper, we introduce and test our algorithm to create a road network representation for city-scale active transportation simulation models. The algorithm relies on open and universal data to ensure applicability for different cities around the world. In addition to the major roads, their geometries and the road attributes typically used in transport modelling (e.g., speed limit, number of lanes, permitted travel modes), the algorithm also captures minor roads usually favoured by pedestrians and cyclists, along with road attributes such as bicycle-specific infrastructure, traffic signals, and road gradient. Furthermore, it simplifies the network's complex geometries and merges parallel roads if applicable to make it suitable for large-scale simulations.
   
   To examine the utility and performance of the algorithm, we used it to create a network representation for Greater Melbourne, Australia and compared the output with a network created using an existing transport simulation toolkit along with another network from an existing city-scale transport model from the Victorian government. Through simulation experiments with these networks, we illustrated that our algorithm achieves a very good balance between simulation accuracy and run-time. For routed trips on our network for walking and cycling it is of comparable accuracy to the common network conversion tools in terms of travel distances of the shortest paths while being more than two times faster when used for simulating different sample sizes. Therefore, our algorithm offers a flexible solution for building accurate and efficient road networks for city-scale active transport models for different cities around the world.
    
\end{abstract}


\section{Introduction}
\label{sec:intro}

Transport models have traditionally been designed for simulating cars and public transportation traffic~\citep{milakis_what_2014}. 
As a result, a common practice in transport models is to only include major roads such as arterial and higher capacity highways, with a limited set of road attributes such as link length, speed limit, and road capacity. 
A vast amount of literature suggests that for walking and cycling, factors such as access to minor roads, availability of footpaths and bicycle lanes, good street connectivity, and attributes of the route such as elevation are also critical ~\citep{buehler_bikeway_2016,prato_evaluation_2018,broach_where_2012,sugiyama2012destination}.
Therefore, given the increasing interest in recent years to also incorporate active modes of transportation such as walking and cycling into the city-scale transportation simulation models~\citep{leao_building_2017,aziz_high_2018,agarwal_bicycle_2019}, there is a strong need for algorithms and tools for producing simulation-ready transport networks fit for active transportation modelling.

In order to incorporate road infrastructure geometries and related metadata, most transport simulation toolkits either provide the capability of reading GIS data directly--often as ESRI shapefiles~\citep{esri1998esri} or
include a tool to convert the GIS files to their required input formats.
However, an input GIS file representing the road network and its required attributes is not always readily available, and where provided by official sources and government departments, it rarely includes the level of details and features needed for modelling walking and cycling.

In recent years, \acrfull{osm}\footnote{\url{https://www.openstreetmap.org}} has become a popular source for acquiring road network geospatial data for active transport research.
\citet{ferster_using_2020} used \osm to extract cycling road infrastructure and to compare it with open data from cities and municipalities in Canada.
\citet{yeboah_route_2015} analysed cyclists' route choice behaviour using \osm and GPS data and found \osm to be a robust data source for extracting transportation network in cycling studies.
Similarly, by comparing \osm to proprietary geodata providers, \citet{zielstra_comparative_2011} found \osm to be a valuable source for pedestrian accessibility studies.

\citet{ziemke_bicycle_2018} developed an algorithm for building a simulation-ready road network for cycling from \osm for \ms~\citep{horni_multi-agent_2016} open-source traffic simulation framework. 
Even though formats such as \osm are tool-agnostic, data processing is often selective and tool-specific, resulting in outputs only fit for consumption by the desired tool or the specific study alone. 
Since extracting a simulation-ready road network from \osm can be quite involved \citep{melnikov_data-driven_2015}, available converters accompanying specific toolkits are often rigid by design.
What is required is a general-purpose and flexible \osm processing workflow that can be configured to achieve the desired balance between network detail (that improves the accuracy of modelling) on the one hand, and network simplification (that improves simulation time) on the other.

\textit{This paper proposes a flexible and general algorithm for building a road network from \osm input for use in city-scale active transportation simulation models.} 
The network generation algorithm (explained in Section~\ref{sec:algorithm}) uses GIS map information for a bounded region extracted from \osm and creates an output network suitable for use in transport simulations. 
The objective was to create an algorithm that balances simplifying the number of nodes and links in the output network for faster routing times in large-scale transport simulations, with capturing sufficient details of minor roads, cycling infrastructure, and footpaths, for an accurate representation of active modes of transportation such as walking and cycling.
The road network output by the algorithm includes the key road data from \osm required for simulating walking and cycling, in addition to cars and public transport.

We expect our algorithm to create networks that are sufficiently detailed to provide the accuracy needed for active transport simulation while minimising the overhead of such a detailed network on simulation run-time.
To examine this, we built a road network for the Greater Melbourne region of Australia using our algorithm and compared it with other networks commonly used in transport modelling in terms of simulation run-time (Section~\ref{sec:simulation}) and travel distances of the shortest paths between random trips (Section~\ref{sec:shortestPath}).
Results of this comparison and how our algorithm can contribute to building simulation models for active transportation are presented in sections~\ref{sec:results} and \ref{sec:discussion}.

\section{Methodology}
\label{sec:method}

\subsection{Network generation algorithm}
\label{sec:algorithm}

\begin{figure}[ht]
    \centering
    \includegraphics[width=\textwidth]{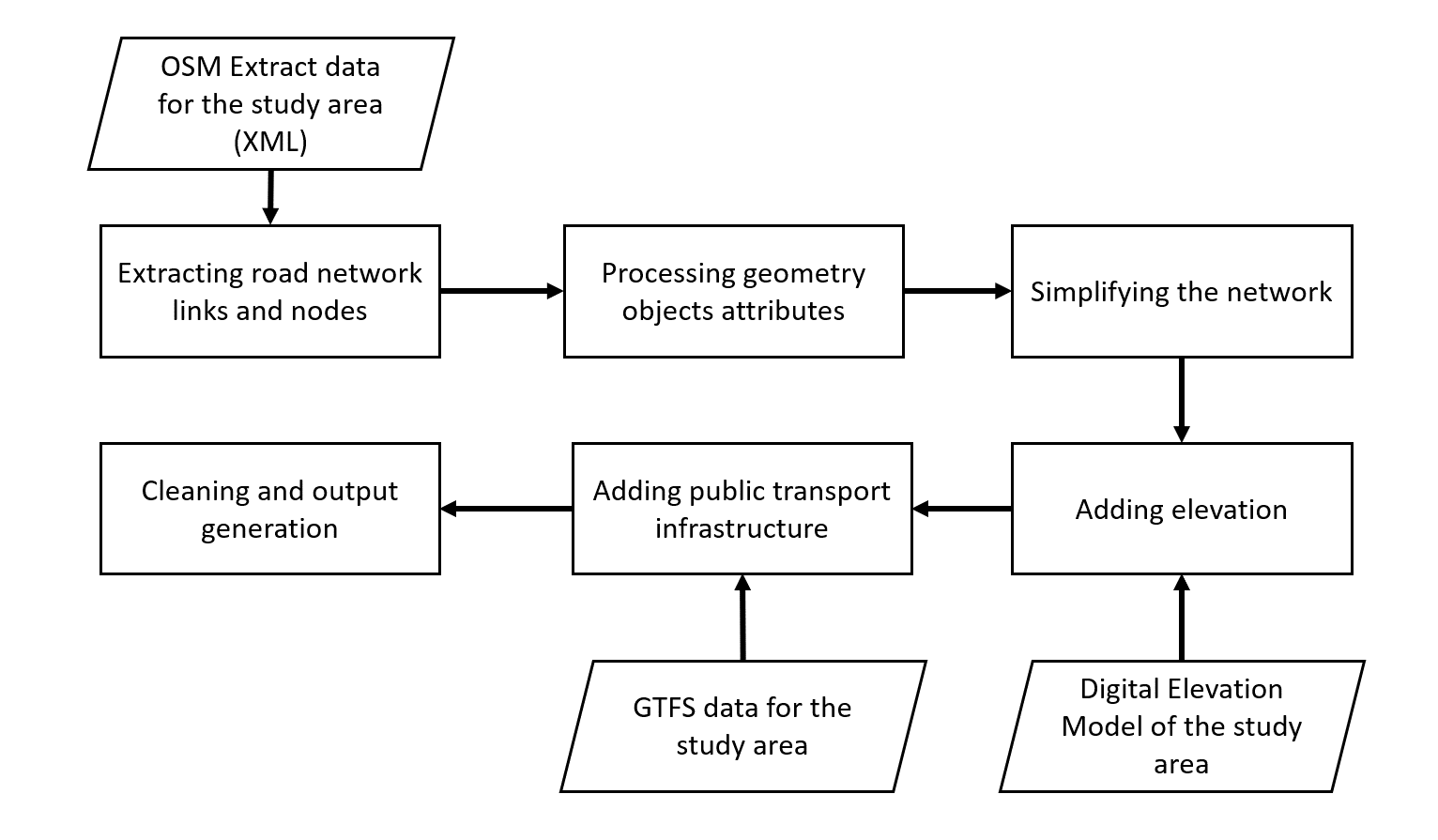}
    \caption{Flow diagram of the network generation algorithm}
    \label{fig:steps}
\end{figure}

Figure~\ref{fig:steps} outlines the six main steps in the network generation process being: 
\begin{inparaenum}[(i)]
\item extraction of the network nodes and links from the \osm input (Section~\ref{sec:step1}); 
\item extraction of their associated \osm key-value attributes such as the number of lanes and speed limit (Section~\ref{sec:step2});
\item simplification of certain geometries like complex roundabouts and intersections (Section~\ref{sec:step3});
\item incorporation of road elevation data necessary for nuanced modelling of cycling patterns (Section~\ref{sec:step4}); 
\item addition of public transport infrastructure (Section~\ref{sec:step5}); and 
\item cleaning to ensure that all included nodes in the network are reachable from every other node to make it simulation-ready (Section~\ref{sec:step6}).
\end{inparaenum}

The algorithm is implemented in the R programming language, with some accompanying functions written in SQL and Bash shell script languages. 
The complete working implementation of the algorithm is available on GitHub\footnote{\url{https://github.com/matsim-melbourne/network}}.

\subsubsection{Extracting road network links and nodes}
\label{sec:step1}

\begin{figure}[ht]
    \centering
    \includegraphics[width=0.8\textwidth]{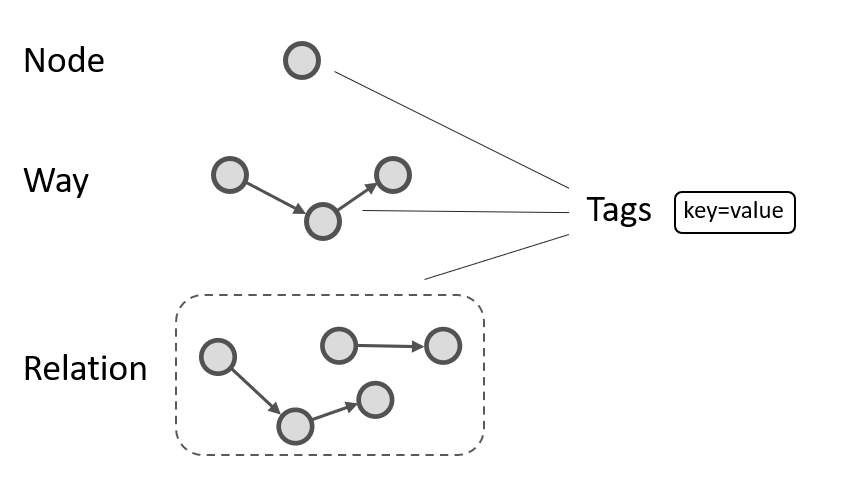}
    \caption{\acrfull{osm} data model}
    \label{fig:osm}
\end{figure}

The first step of the algorithm was extracting the road infrastructure geometries from the raw \osm extract of the study region. 
Figure~\ref{fig:osm} shows the relationship between the elements present in \osm data that form the conceptual model of the real world. 
The simplest element is a \textit{node}, which represents a spatial coordinate. 
Several nodes may be chained together with joining links to form a \textit{way}, which may be open (a polyline) to represent, say, a path, or closed (polygon or area) to represent say an administrative boundary or a lake. 
A \textit{relation}, as the name suggests, captures any natural relationships between nodes and ways (and other relations) for instance where a set of ways might represent a route. Finally, \textit{tags} are key/value pairs that impart meaning to data elements.

In this context, the task of extracting the road network could be summarised as specifying the relevant set of tags that should be used to filter geometries in the input \osm data. 
In general, due to the open nature of \osm data and different tag combinations often being used by volunteer mappers in various locations, a context-specific filtration based on the local tagging practices was needed. 

We used two sources to develop this filter. 
The first was the static tagging guidelines provided to mappers within the region, in our case the Australian Tagging Guidelines\footnote{\url{https://wiki.openstreetmap.org/wiki/Australian_Tagging_Guidelines}}, that provided a starting point for the potential tag-set to use in the filter. 
The second was \osm TagInfo\footnote{\url{https://taginfo.openstreetmap.org}}, which is a dynamic service that regularly analyses \osm data to extract the popularity of tag use. 
Together these sources provided a relevant set of tags that are known to be used to identify relevant infrastructure for different modes of travel in the area of concern.
Using this information, the algorithm first selected elements that contained one of the \texttt{highway} tags given in Table~\ref{tab_osmTagDefs}.
For example, residential streets were extracted using the \texttt{highway=residential} filter.

Network nodes were then added either at the intersection of two links or at the end of a link representing a terminating dead-end street or a cul-de-sac.
Nodes were not added if one of the links is identified as an underpass, tunnel, or bridge, to avoid cases where links seemingly intersected in the 2D plane but in reality, those were disconnected due to a 3D vertical displacement.
Next, information about whether the road intersection contained a traffic light or a roundabout was appended to the node attributes. 
The resulting node and link geometries were exported along with their attributes to an SQLite database for use in the next steps.

\subsubsection{Processing geometry objects attributes}
\label{sec:step2}

Identifying road infrastructure types and extracting relevant road attributes from the \osm link geometries was the second step of the network generation algorithm. 
\osm tag key-value pairs were processed at this step in order to extract required road attributes including the number of lanes, speed limit, permitted modes of transportation, and availability/type of bicycle-specific infrastructure. 
However, since \osm tagging guidelines are not enforced, not all desired information was available for all road segments in \osm. 
Therefore, estimated default values from Table~\ref{tab_osmTagDefs} were used for each road segment wherever \osm information was not sufficient. 

\begin{table}[ht]
	\centering
	{\scriptsize 	
		\caption{\osm road hierarchy and their default values for Australia}
		\label{tab_osmTagDefs}
		\begin{tabular}{@{}lcccccc@{}}
			\toprule
			Road type$^{*}$   & lanes           & Speed limit & Capacity                      & \multicolumn{3}{c}{Permitted modes}  \\ \cmidrule(l){5-7} 
			                & (n)             & (km/h)      & (vehicles per hour, per lane) & Bicycle    & Walk       & Car        \\ \midrule
			Motorway        & 4               &  110        & 2000                          & No         & No         & Yes        \\
			Motorway\_link  & 2               &  80         & 1500                          & No         & No         & Yes        \\
			Trunk           & 3               &  100        & 2000                          & No         & No         & Yes        \\
			Trunk\_link     & 2               &  80         & 1500                          & No         & No         & Yes        \\
			Primary         & 2               &  80         & 1500                          & Yes        & Yes        & Yes        \\
			Primary\_link   & 1               &  60         & 1500                          & Yes        & Yes        & Yes        \\
			Secondary       & 1               &  60         & 1000                          & Yes        & Yes        & Yes        \\
			Secondary\_link & 1               &  60         & 1000                          & Yes        & Yes        & Yes        \\
			Tertiary        & 1               &  50         & 600                           & Yes        & Yes        & Yes        \\
			Tertiary\_link  & 1               &  50         & 600                           & Yes        & Yes        & Yes        \\
			Residential     & 1               &  50         & 600                           & Yes        & Yes        & Yes        \\
			Unclassified    & 1               &  50         & 600                           & Yes        & Yes        & Yes        \\
			Living\_street  & 1               &  40         & 300                           & Yes        & Yes        & Yes        \\
			Cycleway        & 1               &  30         & 300                           & Yes        & No         & No         \\
			Track           & 1               &  30         & 300                           & Yes        & No         & No         \\
			Service         & 1               &  40         & 200                           & Yes        & Yes        & Yes        \\
			Pedestrian      & 1               &  30         & 120                           & No         & Yes        & No         \\
			Footway         & 1               &  15         & 120                           & No         & Yes        & No         \\
			Path            & 1               &  15         & 120                           & No         & Yes        & No         \\
			Corridor        & 1               &  15         & 50                            & No         & Yes        & No         \\
			Steps           & 1               &  15         & 10                            & No         & Yes        & No         \\ \bottomrule
			\multicolumn{7}{l}{\scriptsize{* Based on OSM standard, might be different to the road categories commonly used in different cities.}} \\
	\end{tabular}}
\end{table} 

Furthermore, a number of \osm tag combinations were used to identify different types of bicycle infrastructure including: 
\begin{itemize}
    \item Bike paths, i.e., off-street infrastructure dedicated to cyclists -- common \osm tag combination: \texttt{Highway=cycleway AND foot=no}; 
    \item Shared paths, i.e., off-road paths shared between cyclists and pedestrians -- common \osm tag combination: \texttt{Highway=cycleway AND foot=yes}; 
    \item Simple bike lanes, i.e., on-street bike lanes that are marked on the street but not separated via a physical barrier -- common \osm tag combination: \texttt{Highway!=cycleway AND cycleway=lane}; 
    \item Separated bike lanes, i.e., on-street lane that are separated from motorised traffic via a physical barrier -- common \osm tag combination: \texttt{Highway=cycleway AND cycleway=track}; 
    \item Shared streets, i.e., streets without a specific bike lane but with priority given to cyclists -- common \osm tag combination: \texttt{Highway!=cycleway AND cycleway=shared\_lane}; and 
    \item No infrastructure, i.e., streets where cycling is allowed but with no bicycle-specific infrastructure -- common \osm tag combination: \texttt{Highway!=cycleway AND cycleway=NA}; 
\end{itemize}

\subsubsection{Simplifying the network}
\label{sec:step3}

\begin{figure}
    \centering
    \begin{subfigure}[b]{0.44\textwidth}
        \includegraphics[width=\textwidth]{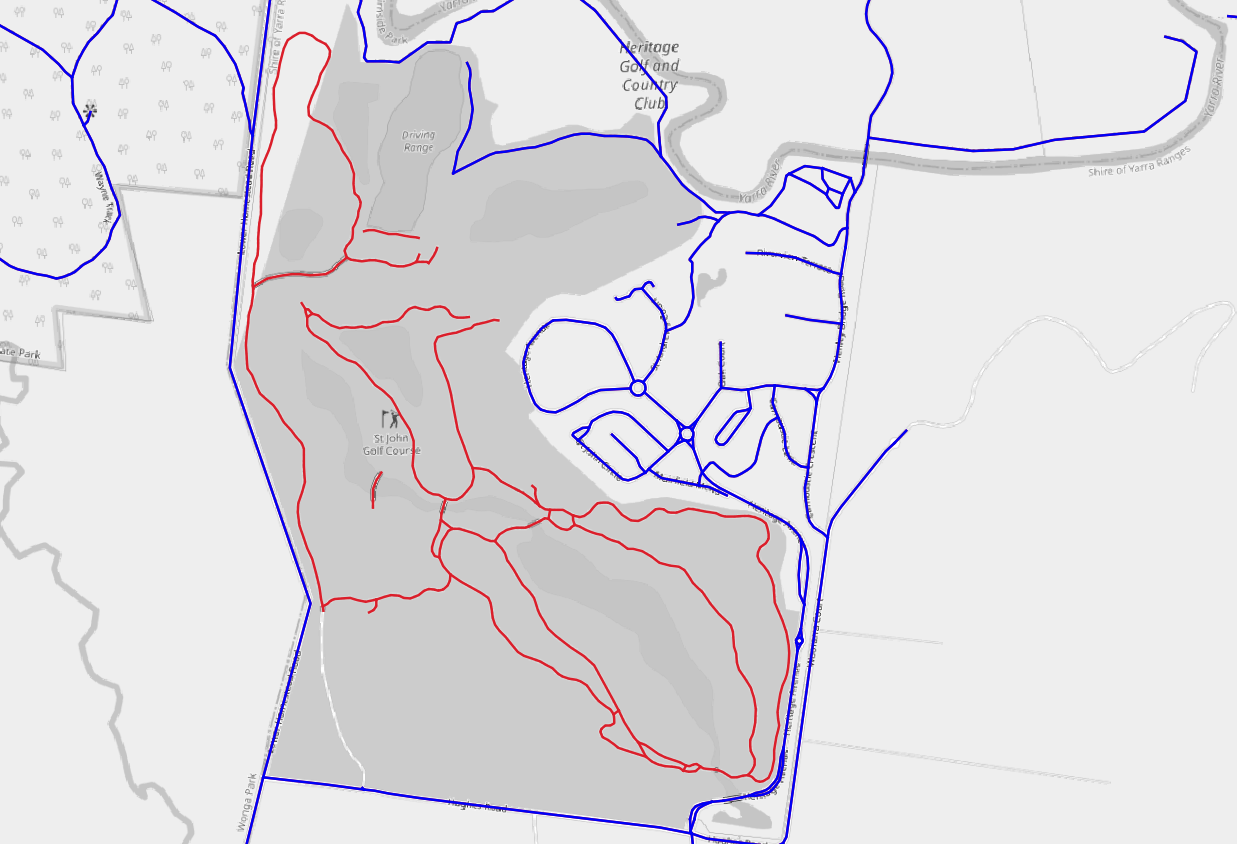} 
 		\caption{A golf course}
 		\label{fig:disconnectedLinksGC}
    \end{subfigure}
    \begin{subfigure}[b]{0.44\textwidth}
        \includegraphics[width=\textwidth]{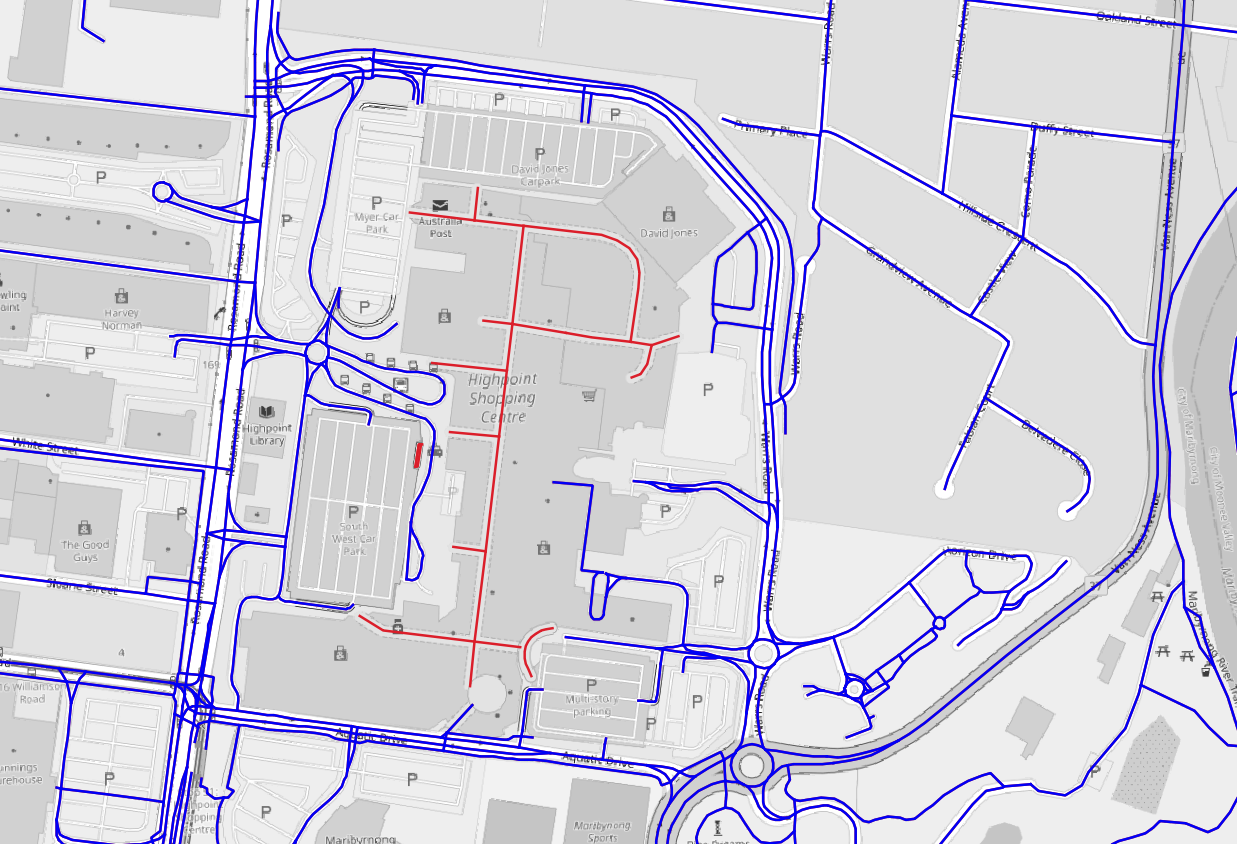} 
 		\caption{A shopping centre}
 		\label{fig:disconnectedLinksSC}
    \end{subfigure}
    \caption{Disconnected links within enclosed area (links in red) to be removed by the algorithm (base map from \textcopyright\href{https://www.openstreetmap.org/copyright}{OpenStreetMap}) }
    \label{fig:disconnectedLinks}
\end{figure}

At this point of the algorithm, the augmented network with road attributes still encompasses small disconnected parts, meaning that the network has islands of link clusters that were unreachable from other parts of the network.
These components were usually links within enclosed areas disconnected from road facilities such as within golf courses or shopping centres that were not yet filtered out (Figure~\ref{fig:disconnectedLinks}).
To resolve this, the largest connected cluster of links was selected by the algorithm and the rest was discarded. 
This yielded a fully connected network in which it was possible to reach any destination from any starting location, ignoring travel modes.

 \begin{figure}[ht]
 	\centering
 	\begin{subfigure}[b]{0.44\textwidth}
 		\includegraphics[width=\textwidth]{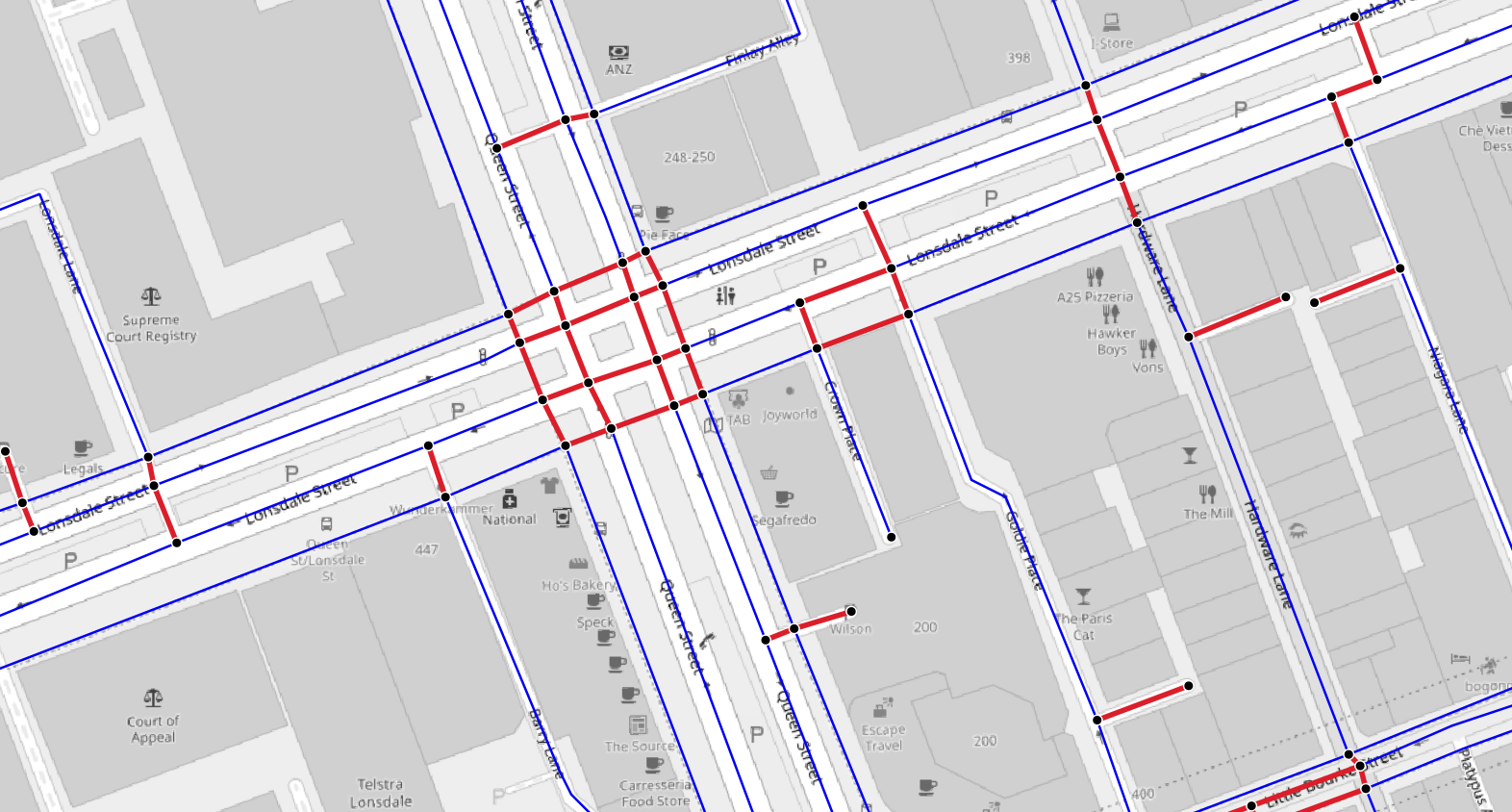} 
 		\caption{Intersection - no simplification}
 		\label{fig:intersectionsBefore}
 	\end{subfigure}
 	\begin{subfigure}[b]{0.44\textwidth}
 		\includegraphics[width=\textwidth]{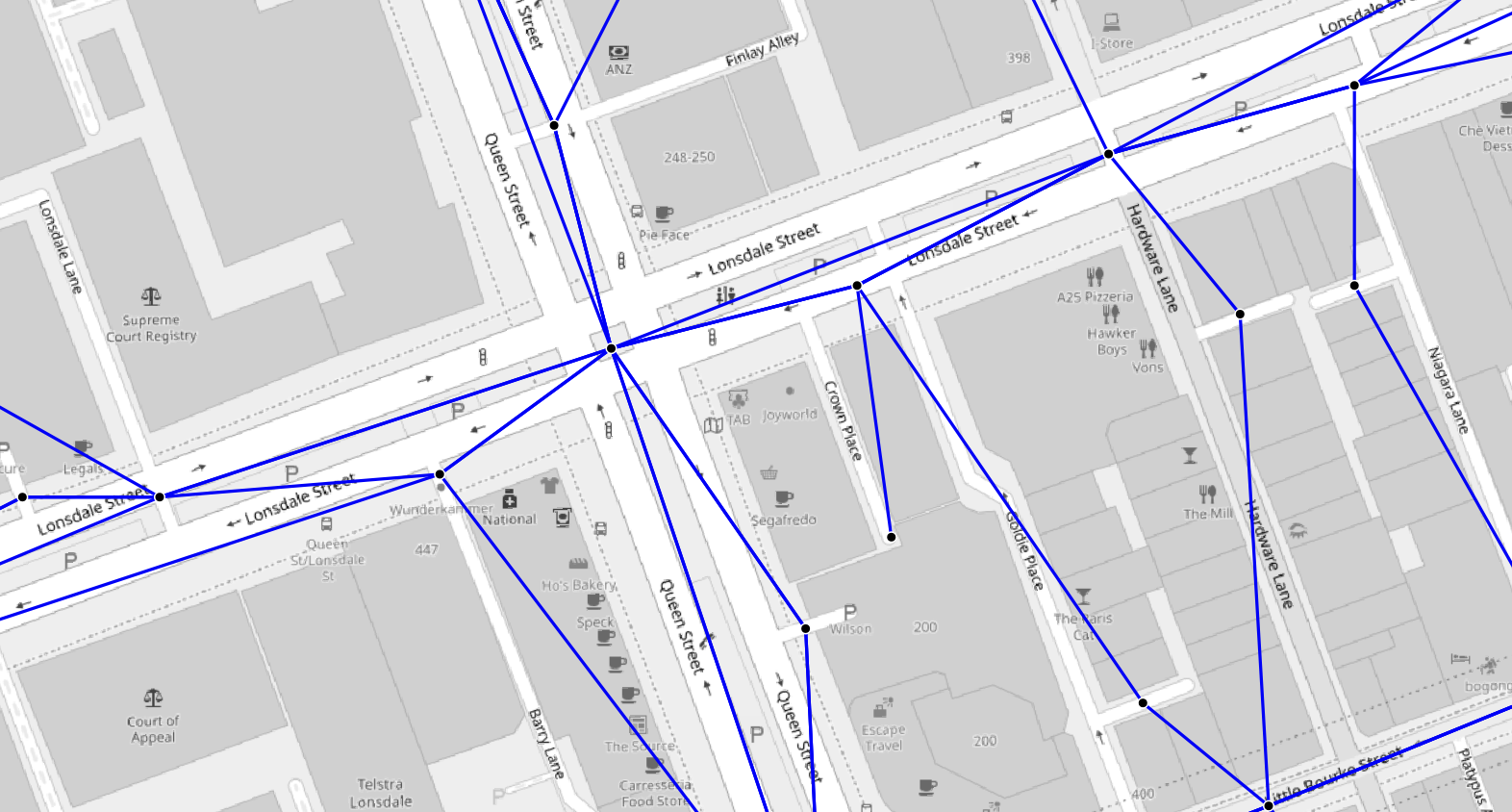} 
 		\caption{Intersection - with simplification}
 		\label{fig:intersectionAfter}
 	\end{subfigure}
 	
 	\begin{subfigure}[b]{0.44\textwidth}
 		\includegraphics[width=\textwidth]{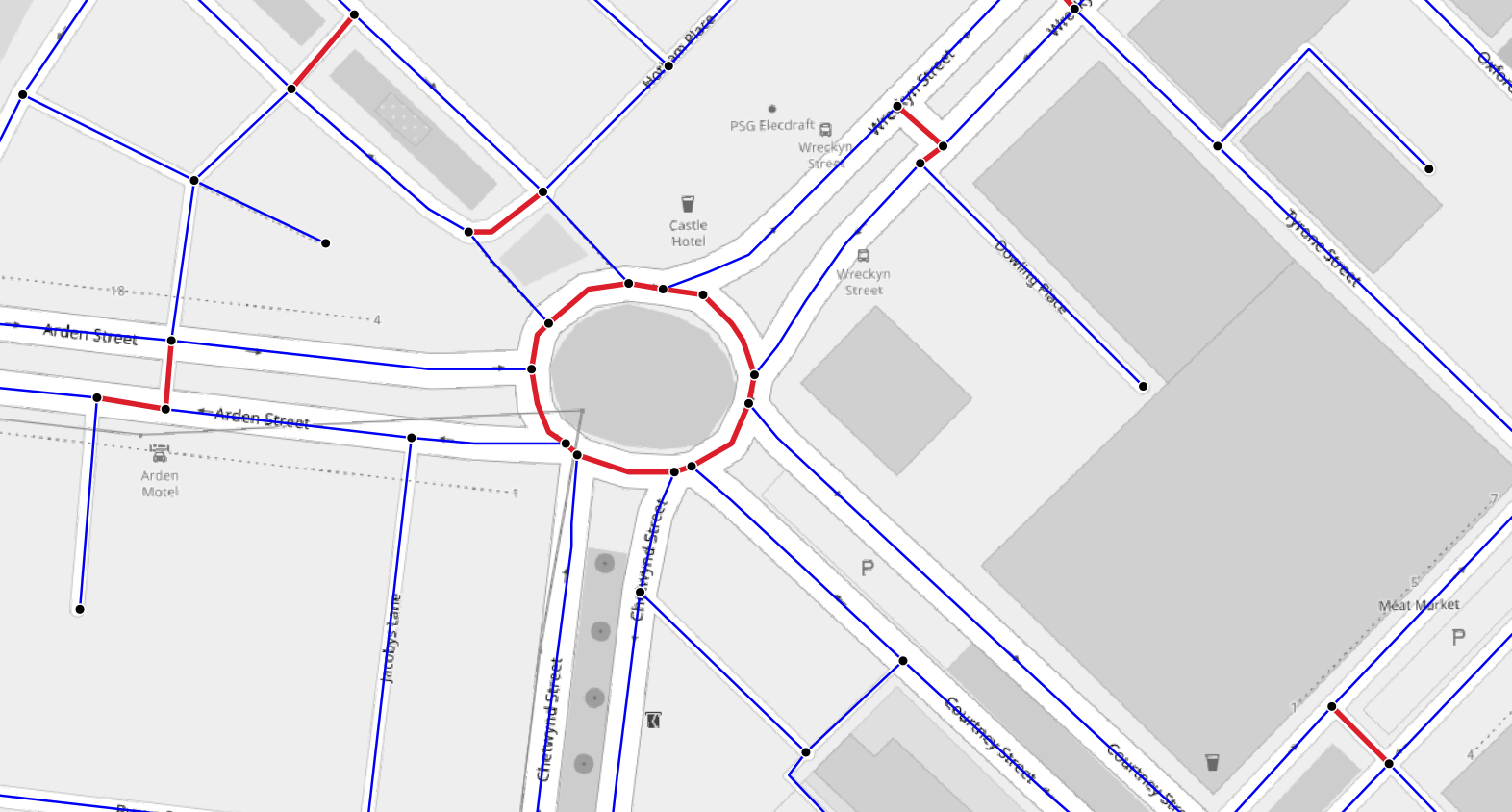}
 		\caption{Roundabout - no simplification}
 		\label{fig:roundaboutBefore}
 	\end{subfigure}
 	\begin{subfigure}[b]{0.44\textwidth}
 		\includegraphics[width=\textwidth]{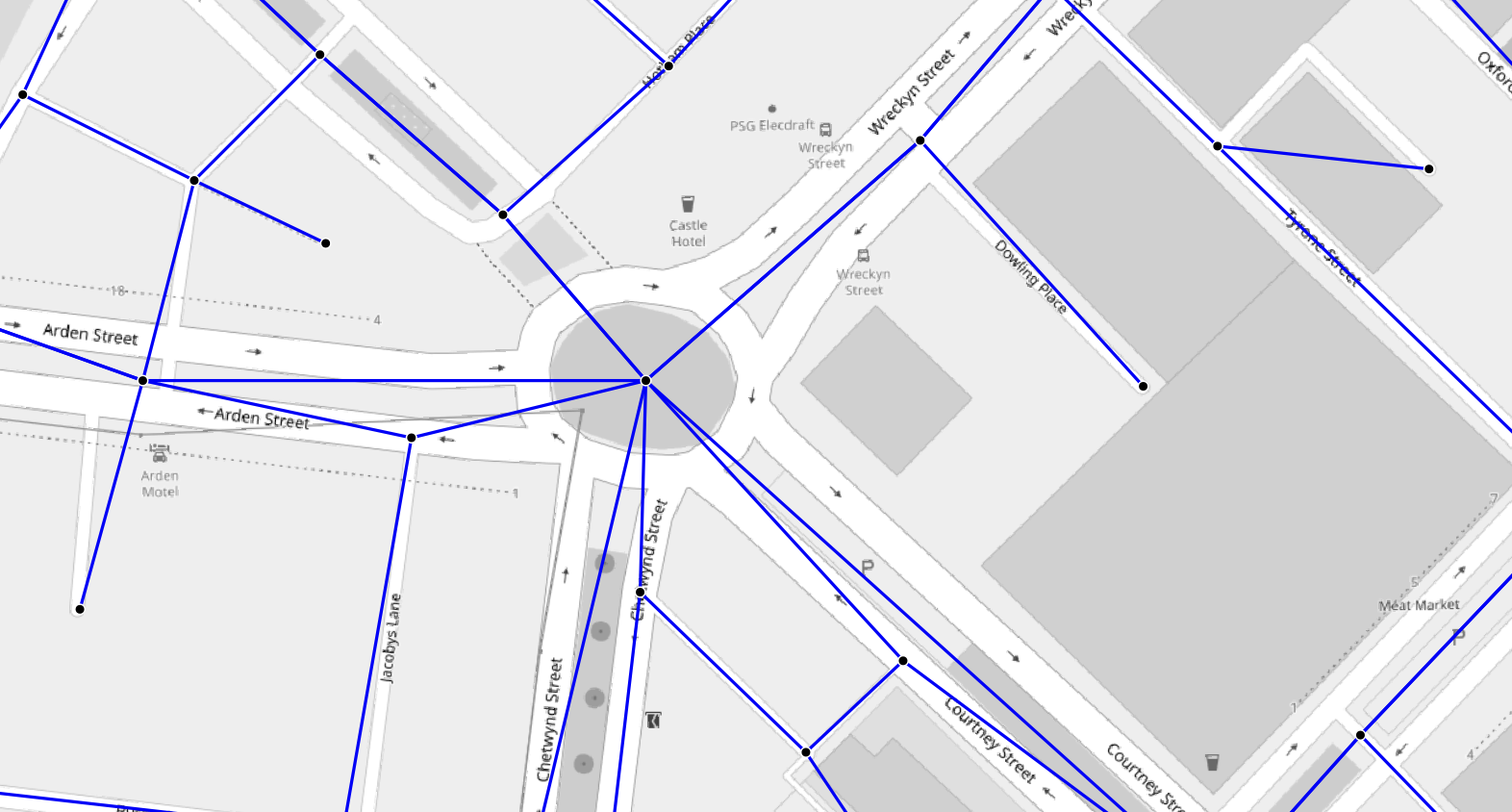} 
 		\caption{Roundabout - with simplification}
 		\label{fig:rounadbouthAfter}
 	\end{subfigure}
 	
 	\caption{Intersections and roundabouts before and after network simplification, with a minimum link length threshold of 20m, links shorter than the threshold are highlighted in red, (base map from \textcopyright\href{https://www.openstreetmap.org/copyright}{OpenStreetMap})}
 	\label{fig:simplification}
 \end{figure}

Complex intersections and roundabouts in the network up to this point were represented using several small links as illustrated in Figures~\ref{fig:intersectionsBefore} and \ref{fig:roundaboutBefore}.
There might be modelling use-cases where parallel links and complex intersections and roundabouts are important, for instance for understanding pedestrian movements on a given side of the road. 
In most cases, however, these complex geometries (Figures~\ref{fig:intersectionsBefore} and \ref{fig:roundaboutBefore}) do not add significant benefits to the accuracy of the city-scale models and could be simplified. 
To do this, the algorithm merged all cases of two neighbouring nodes connected with a common link and with a distance less than a \textit{minimum link length} threshold of 20m.
As a result, small roundabouts and complex intersections were reduced to a single node (Figures~\ref{fig:intersectionAfter} and \ref{fig:rounadbouthAfter}). 

Merging nodes in this way resulted in some links forming loops (where a link has the same start and end node) and other sets forming parallel links (where multiple links were assigned the same start and end nodes).
Loops were simply removed, and parallel car-only links were merged by selecting the shortest geometry and aggregating attributes from all the links into one. 
Specifically, one-way links going in the same direction were merged into a single one-way link in a similar direction to the original links.
Two-way links were also merged into a single two-way link.
Then one-way links in opposite directions were merged into a single two-way link.
Any remaining chains of links between intersections were also then merged together, taking the sum of their lengths and the maximum of the link attributes (i.e., speed limit, number of lanes, capacity). 

If a link was less than 500m long, and one of its nodes did not join any other links (a dead-end for example), it was removed by the algorithm. 
This had the effect of removing minor end-of-trip links such as courts and cul-de-sacs, allowing for further link simplification. 
Therefore, the link merging processes explained above were run one more time. 
Next, road geometries were simplified into straight lines.
We expect converting to straight lines not to have major implications for the accuracy of the network as road length was already calculated and saved with the link attributes in previous steps, making complex geometries redundant. 

A possible side-effect of the merging process of this step was that an off-road bike path close to the main road to be merged with that road.  
Although for most cases this should not be problematic, there are scenarios where it might be important to keep them separated, for example, if car-bicycle interactions are of interest.
Therefore, at the end of the simplification step, any off-road bike paths merged with car traffic were separated to ensure the physical separation from motorised traffic was also maintained in the output network.  

\subsubsection{Adding elevation}
\label{sec:step4}

Road elevation is particularly useful for modelling cycling transportation.
For adding road elevation, a 10m LiDAR-derived \gls{dem} from the Victorian Department of Environment, Land, Water and Planning\footnote{The Digital Elevation Model (DEM) 10m grid of Victoria (bare earth) derived from LiDAR model, available at \url{https://discover.data.vic.gov.au/dataset/vicmap-elevation-dem-10m}} was used. 
Since the addition of road elevation relied on external information other than \osm that might not be available in all locations, this step was considered to be an optional step that could be bypassed in the full run of the algorithm, 

Bilinear interpolation was used to determine the height of each node in the network.
Bilinear interpolation uses the weighted average of the four closest cell centres to calculate the values of a grid location and is shown to be a simple but adequate method comparing to more complex approaches such as kriging \citep{rees_accuracy_2000}.
Weights in bilinear interpolation were based on the distance of the cell centres, i.e., the further away a cell was the less influence it had.

\subsubsection{Adding public transport infrastructure}
\label{sec:step5}

Similar to the previous step, addition of \gls{pt} infrastructure was also considered an optional step as it requires extra inputs.
The \gls{pt} network was constructed based on \gls{gtfs} data \citep{google2020gtfs}\footnote{The \gls{gtfs} data was sourced from Victoria's state transit agency in September 2019, \url{https://www.ptv.vic.gov.au/footer/data-and-reporting/datasets/}}.
\gls{pt} stations and weekday vehicle schedules from the \gls{gtfs} feed were extracted by the algorithm.
To ensure overall network connectivity, \gls{pt} stations were snapped to the nearest network nodes that were accessible via walking, cycling, and driving. 
Direct links were then added between station nodes along each transit route to represent the \gls{pt} network links.
A list of \gls{pt} vehicles and their schedules were also extracted from \gls{gtfs} and written into XML format to be used in \gls{pt} modelling.

\subsubsection{Cleaning and output generation}
\label{sec:step6}

Finally, the largest connected component of the network for each mode was selected to ensure every node of the network is reachable from all other nodes using desired travel modes. 
This step is similar to that described in Section~\ref{sec:step3} but with the main difference that here directed connectivity was determined separately for each travel mode, whereas in the former step modes were not considered.
Default travel modes included in this process were walk, bicycle and car. 

Lastly, the final reachable network was converted into different output formats to be used for various purposes.
These output formats include: 1) an ESRI shapefile to be used as the input in transport models, 2) a SpatiaLite database that can be used for further analysis, and 3) an XML file to be used for the simulations of this paper explained in Section~\ref{sec:simulation}.

\subsection{Network shortest path comparison}
\label{sec:shortestPath}

\begin{figure}[h]
    \centering
    \includegraphics[width=0.7\textwidth]{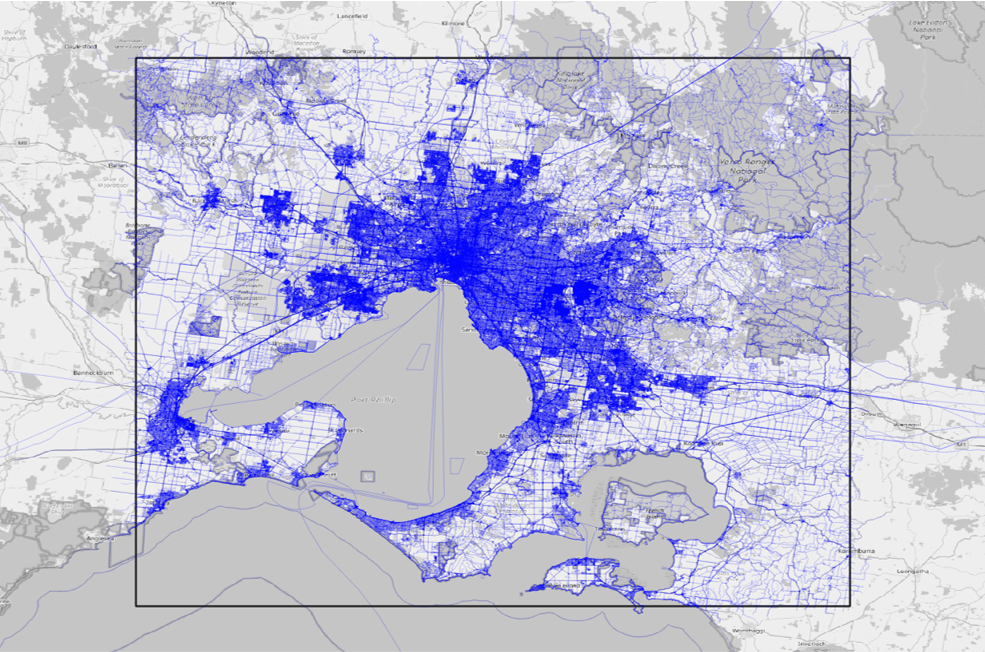}    
    \caption{Raw \osm extract of the study area (roads layer) for Greater Melbourne. (base map from \textcopyright\href{https://www.openstreetmap.org/copyright}{OpenStreetMap} and the bounding box for Greater Melbourne from \url{https://github.com/interline-io/osm-extracts/})}
    \label{fig:GMelb}
\end{figure}

The algorithm explained in Section~\ref{sec:algorithm} was used to create a network for Greater Melbourne in Australia--a region covering approximately 10,000 km$^2$ and housing Melbourne's population of roughly 5 million residents.
Figure~\ref{fig:GMelb} illustrates the study area and the \osm extract used by the algorithm as its input. 
The output network was then analysed to better understand its accuracy and usefulness when compared with other common networks used in transport modelling. 
Travel distance of the shortest paths for a sample of walking, cycling and driving trips were considered as measures to analyse accuracy of the algorithm's output .
\gls{pt} was not included as it typically has its own dedicated infrastructure and follows a predetermined route and schedule rather than the shortest paths. 

\subsubsection{Road networks for shortest path comparison}
\label{sec:networksGen}

The three road networks that were compared with respect to travel distances using different modes are briefly described here.

\begin{figure}[ht]
\centering
{\scriptsize
 	\begin{subfigure}[b]{0.32\textwidth}
 		\includegraphics[width=\textwidth]{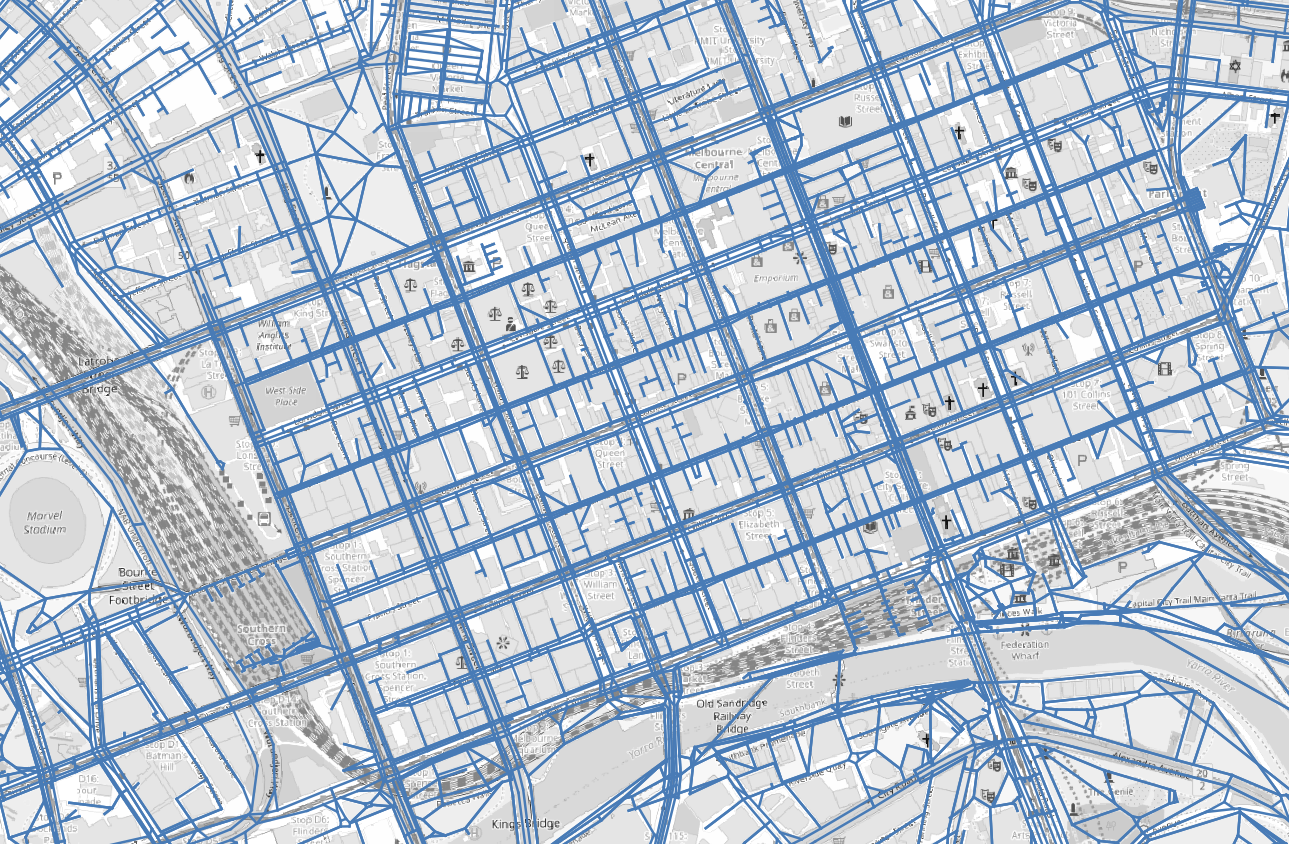} 
 		\caption{\acrshort{mn}}
 		\label{fig:osmInnMelb}
 	\end{subfigure}
 	\begin{subfigure}[b]{0.32\textwidth}
 		\includegraphics[width=\textwidth]{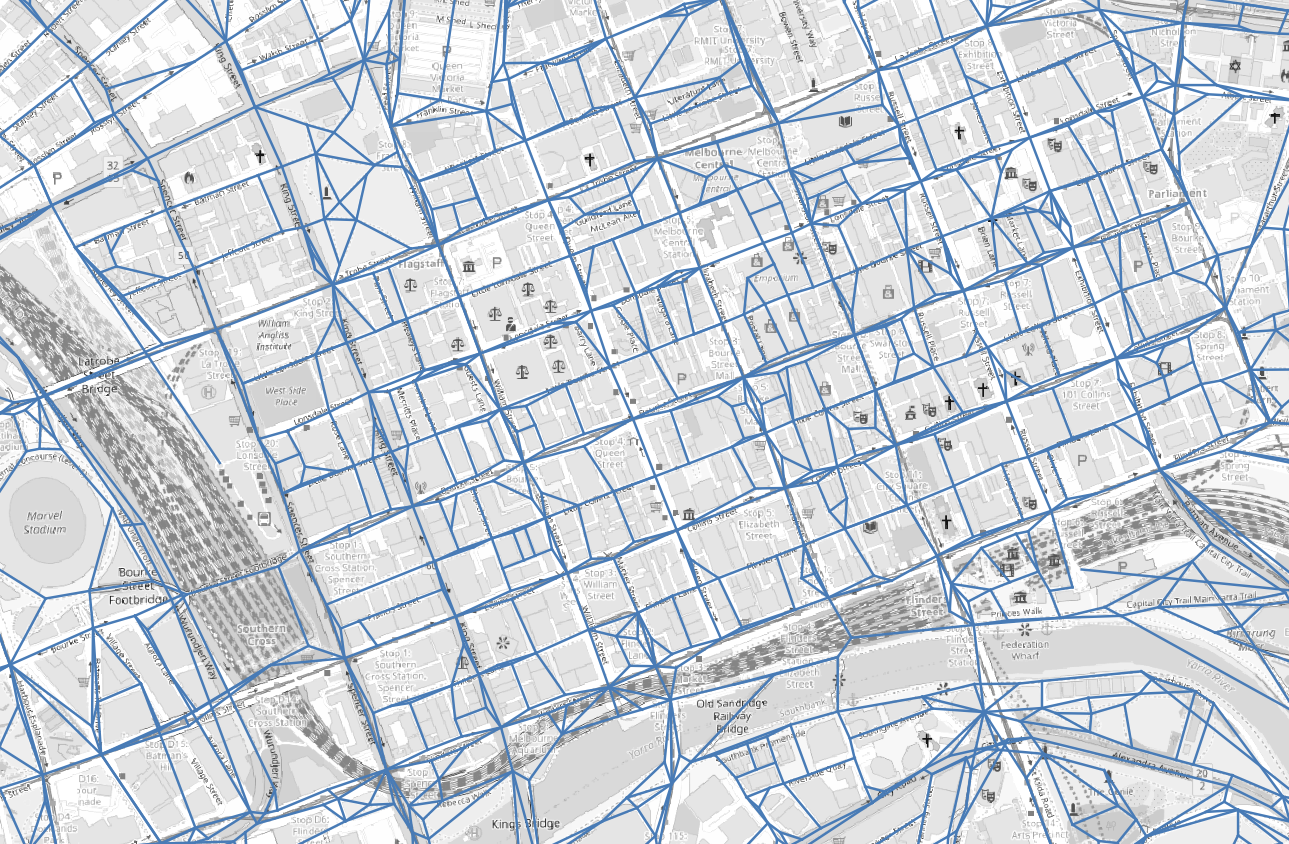}
 		\caption{\acrshort{sn}}
 		\label{fig:allLinksInnMelb}
 	\end{subfigure}
 	\begin{subfigure}[b]{0.32\textwidth}
 		\includegraphics[width=\textwidth]{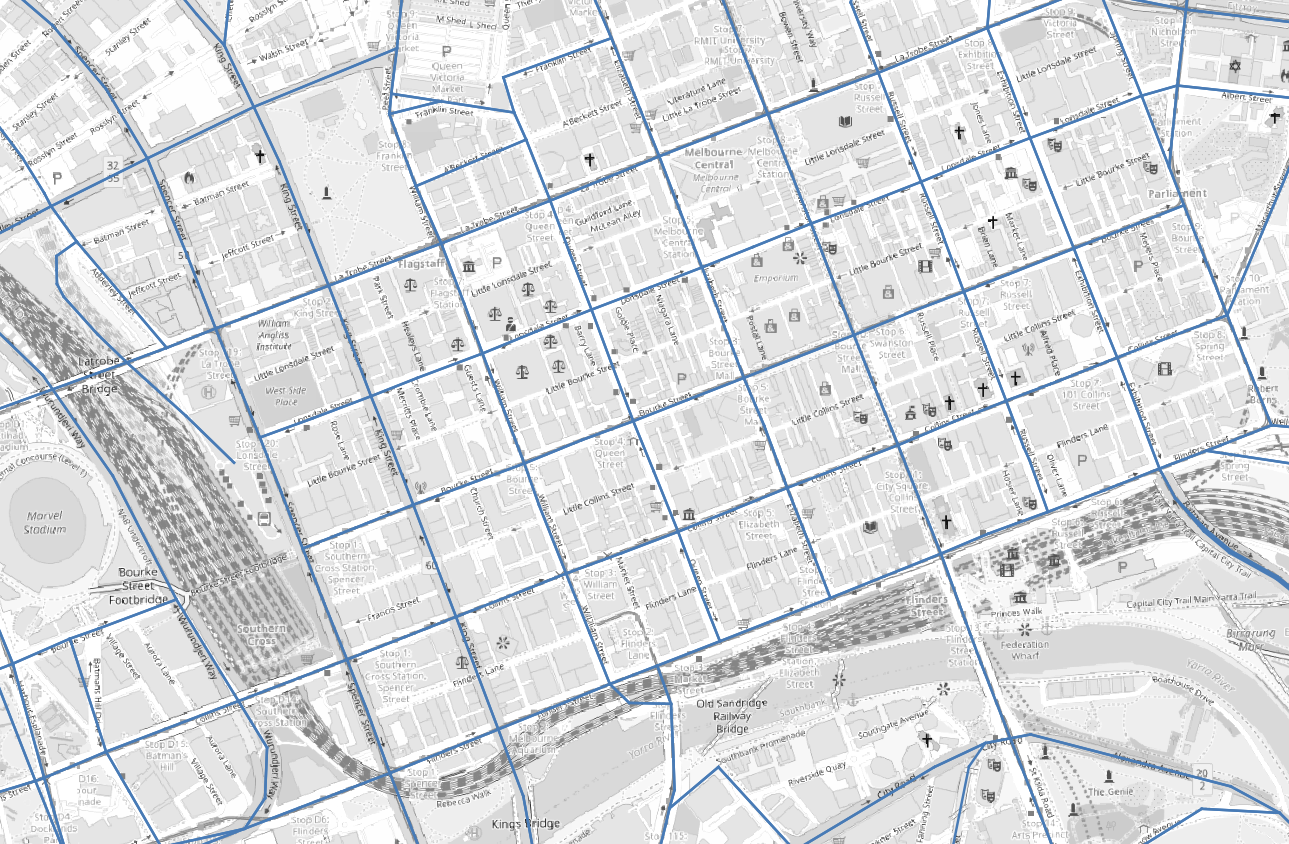}
 		\caption{\acrshort{ivn}}
 		\label{fig:majorLinksInnMelb}
 	\end{subfigure}
 	}
 	\caption{Melbourne's Central Business District area represented in \acrfull{mn}, \acrfull{sn}, and \acrfull{ivn}, (base map from \textcopyright\href{https://www.openstreetmap.org/copyright}{OpenStreetMap})}
 	\label{fig:networks}
 \end{figure}

\paragraph{\acrfull{mn}} The first network was an exact \osm conversion, representing networks created through existing network generation utilities provided in common transport modelling software and toolkits.
This non-simplified \osm network was built using the \ms network conversion extension and was adjusted to incorporate road types from Table~\ref{tab_osmTagDefs} \citep{horni_multi-agent_2016}.
The \mn for the study area consisted of 1,178,167 links and 482,583 nodes (Figure~\ref{fig:osmInnMelb}).

\paragraph{\acrfull{sn}} The second network that we examined was built using our algorithm (Section~\ref{sec:algorithm}) with default settings including a minimum link length of 20m.
Similar to \mn, this network also covers all road types listed in Table \ref{tab_osmTagDefs}.
The \sn was made up of 242,192 links and 144,660 nodes (Figure~\ref{fig:allLinksInnMelb}).

\paragraph{\acrfull{ivn}} The third network represents the types of networks common in strategic transport models for car and public transport simulation.
These types of networks usually lack minor roads and attributes needed for modelling walking and cycling, however, as they are already being used for transport modelling, they could be sometimes seen as the most straightforward option for using in developing an active transportation simulation model as well.
Therefore, we included these types of networks in our comparisons to examine if they are sufficient for modelling walking and cycling or more detailed networks are needed. 
We used the network from Victoria's state of the art activity-based transport model \citep{infrastructure_victoria_model_2017} as the third network.
\citet{infrastructure_victoria_model_2017} model was only used for modelling car and public transport traffic, and not active transportation, therefore this network consists of roads approximately at \texttt{tertiary\_link} level and higher in Table~\ref{tab_osmTagDefs} in addition to public transport railways.
We filtered out the railways given we excluded them in the two other networks as well.
The \ivn was the smallest network consisting 60,973 links and 33,540 nodes (Figure~\ref{fig:majorLinksInnMelb}).

\subsubsection{Building the random sample trips for shortest path comparison}
\label{sec:rndTripsGen}

As illustrated in Figure~\ref{fig:networks}, the level of detail was different in each network so it was expected that a node selected in \mn may not have a counterpart in \ivn. 
The comparison, however, required that any given trip to have same origin and destination points in all three networks.
To achieve this, we first filtered nodes from the smallest network, being \ivn in this case, to a subset of its nodes that were accessible via all three modes of walk, bicycle, and car.
Then, this subset was further filtered to the nodes that had counterparts in the other two networks within a 10m radius.
We refer to the set of remaining nodes from \ivn as the \textit{routable nodes set}, or \n.

\begin{figure}[ht]
 	\centering
 	\includegraphics[width=0.95\textwidth]{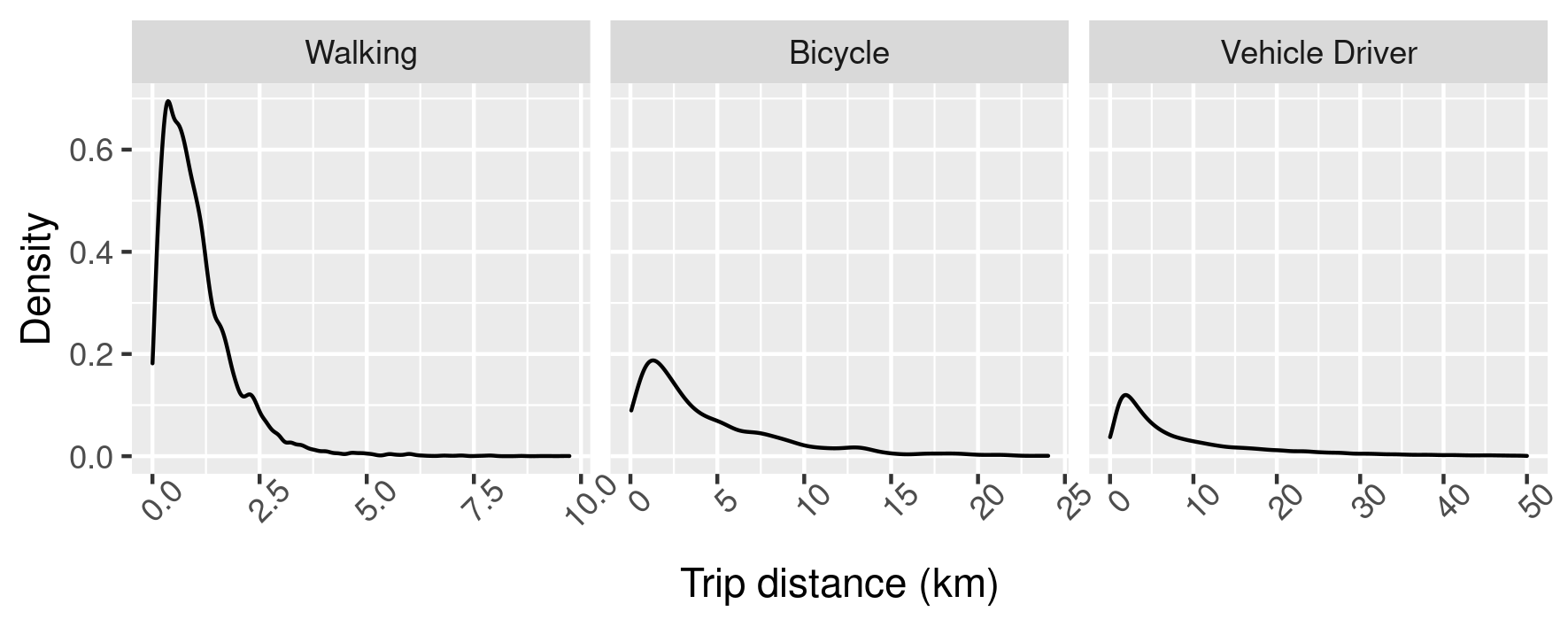}
    \caption{Trip distance density plot for different travel modes based on VISTA 2012-16}
 	\label{fig:vistaDist}
 \end{figure}

1,000 nodes from \n were then randomly selected as the origins of sample trips.
Actual trip distance distributions from \gls{vista} 2012-16~\citep{department_of_transport_victorian_2016} for each travel mode were used to find destination nodes within plausible distances of the origin, given the mode of transportation (Figure~\ref{fig:vistaDist}).
To do this, a log-normal distribution was first fitted to \gls{vista} trip distances for each mode. 
Then, for each origin and travel mode, one destination point was selected from \n based on its fitted log-normal distance distribution, which was no further than $\mu + 2\times \sigma$ distance of the origin.
The nearest node geometries in \mn and \sn to the selected origins and destinations from \ivn were identified and used to build the sample trips in those networks as well.

\subsubsection{Shortest paths estimation and comparison}
\label{sec:spComparison}

The shortest path between the origin and destination nodes for the sample trips on all three networks and travel modes were then compared with each other to examine the accuracy of the network generation algorithm.
The shortest path calculation was done using Dijkstra's algorithm and was implemented using \textit{igraph} package in R.

\mn was considered as the baseline network for the comparison given it was the most detailed of the three, and therefore expected to be closest to real-world infrastructure. 
Pair-wise difference in travel distance for each trip for the given mode when occurring on \ivn and \sn was then compared with their counterparts on \mn.
The error percentage for each trip, $\epsilon$, on \sn and \ivn was considered to be average of absolute values of travel distances difference relative to the same trip on \mn.
As an example, for a given walking trip, $i$, on the \sn, $\epsilon$ was calculated as:

\begin{equation}
    \label{eq:theta}
    \epsilon_{i,\sn}^{walk}=  \frac{|d_{i,\sn}^{walk} - d_{i,\mn}^{walk}|}{d_{i,\mn}^{walk}} \times 100 ,
\end{equation}

where $d_{i,\sn}^{walk}$ represents the shortest path distance of trip $i$ on \sn.
The average of the error percentages for all trips with a given mode $m$, $E_{m}$, was then calculated for aggregate level comparisons.

\subsection{Simulation run-time comparison}
\label{sec:simulation}

In addition to the shortest path comparison explained in Section~\ref{sec:shortestPath}, a series of simulations were also conducted to examine how the output network (\sn) performs in a transport simulation model, when measured for efficiency (simulation run-time), compared with the other networks (\mn and \ivn). 
For compatibility with the rest of the paper, walking, cycling and driving were the main modes of transportation included in the simulation model for the run-time comparison.
For trips using other modes, such as public transport trips, agents were set to take the beeline route from origin to destination and disregard the network roads. 

\subsubsection{Simulation framework}
\label{sec:matsim}

\ms was selected as the simulation toolkit for this exercise \citep{horni_multi-agent_2016}. 
\ms is an activity-based and agent-based transport modelling toolkit that simulates a single day of the transportation system.
In \ms, each agent follows a travel \textit{plan}, listing the locations and timings of each activity that the agent should do during the simulation day, and how it should travel to those locations. 
An initial daily plan for all agents in the population, along with a road network that agents should travel on, are the two main inputs of a \ms model.

Determination of the likely traffic patterns on the given transport infrastructure (the supply) given the daily travel plans of the population (the demand) in \ms is done using a co-evolutionary algorithm. Here each agent tries to find its optimal plan (i.e. one that generates the maximum utility) by trying different possibilities through a number of iterations. A stable solution is then deemed as a quasi Nash-equilibrium where no agent can be better off than it is, given the choices of all the other agents. 

MATSim's co-evolutionary algorithm can be explained as a loop, the \textit{\ms loop}, that gets repeated at each iteration and has three main components: 
\begin{inparaenum}[(i)]
\item mobility simulation or \textit{mobsim}, where all selected travel plans get simulated simultaneously;
\item \textit{scoring}, where every agent's travel experience is scored using a shared utility function; and
\item \textit{re-planning}, where agents are allowed to modify their plans for the next iteration, typically through re-routing, mode shift or changes in the timings of activities~\citep{horni_multi-agent_2016}.
\end{inparaenum}

\subsubsection{Generating agents and their daily plans}
\label{sec:demandGen}

A subset of \gls{vista} 2012-16 data~\citep{department_of_transport_victorian_2016}  was used to create the list of agents and their travel plans. 
\gls{vista} 2012-16 is a sample of 45,562 individuals in Victoria, Australia, and their travel diaries for the survey day. 
\gls{vista} records were filtered to only those with at least one trip on the survey day and living inside the study area, i.e., Greater Melbourne, resulting in a subset of 34,361 individuals.
These agents had a total of 120,213 trips, of which 90,388 were car-based (either as a driver or passenger), 1,973 were bicycle trips, 18,897 were walking, and the remainder were a combination of \gls{pt} and other modes such as taxi or school bus.

Using the resulting subsets and using the sampling weights for average weekday provided in \gls{vista}, three samples of 35,178 agents, 175,893 agents and 351,780 agents were created, representing approximately 1\%, 5\% and 10\% of the travelling population on an average weekday in Greater Melbourne. 
The finest level of geographical location detail in \gls{vista} 2012-16 in Australian Bureau of Statistics standard was at the Statistical Area Level 1 (SA1), an area with an average population of 400 people.
The centroid of the SA1 region for each stop of a \gls{vista} trip was taken as the location of that stop.
These sample populations and travel plans were then converted to a \ms readable XML format to be used as the input for the simulation\footnote{Code is available at \url{https://github.com/jafshin/matsim-melbourne-example/tree/master/population}}. 

\subsubsection{Simulation setting} 
\label{sec:setting}

To examine the performance of the different networks, each network (\sn, \mn, and \ivn) was used to simulate the three sample sizes described in Section~\ref{sec:demandGen} using \ms.
Simulation experiments (i.e., each network and population combination) were run for 50 iterations and the average iteration run-time was recorded for comparison.
Driving, walking and cycling were considered to be the main modes in these experiments and the \ms scoring parameters suggested by \citep{ziemke_bicycle_2018} were used as model parameters.

\section{Results}
\label{sec:results}

\begin{figure}[h!]
    \centering
	\begin{subfigure}[b]{0.45\textwidth}
        \includegraphics[width=\textwidth]{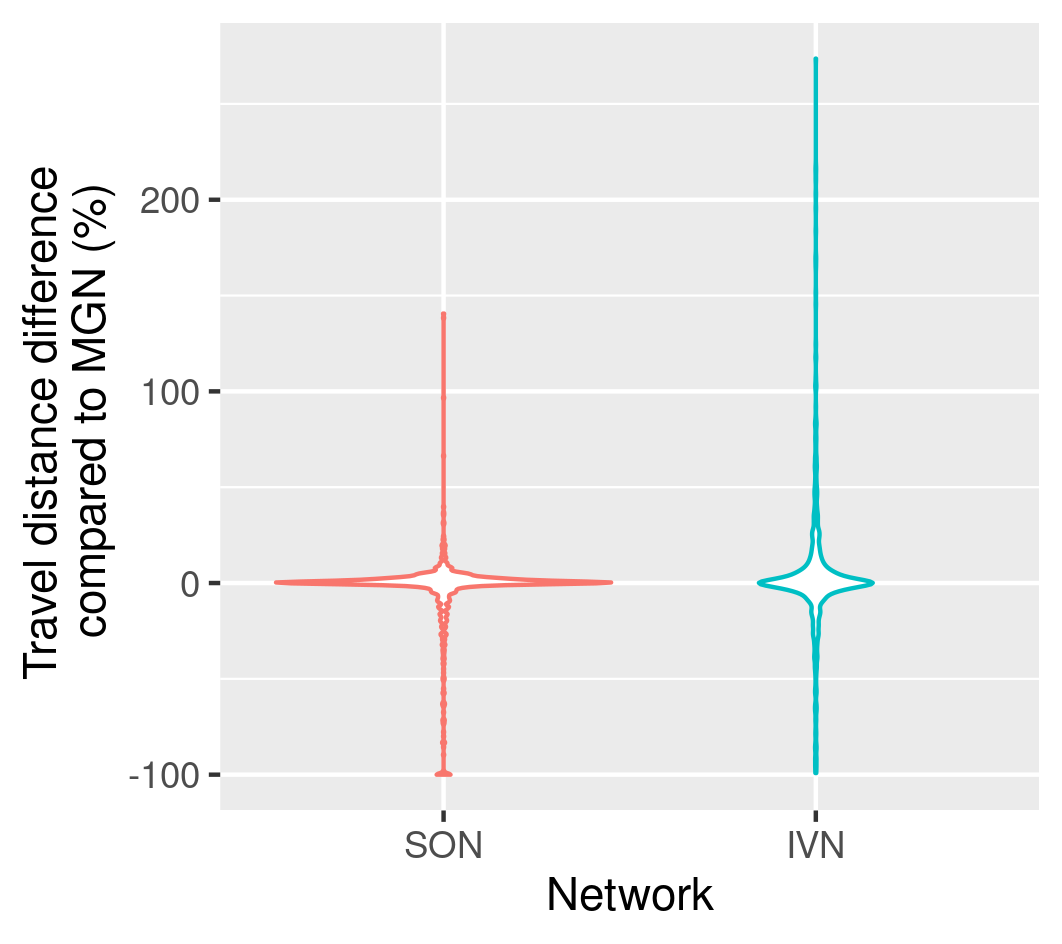}
        \caption{Walking}
        \label{fig:distWalk}
	\end{subfigure}
    \begin{subfigure}[b]{0.45\textwidth}
        \includegraphics[width=\textwidth]{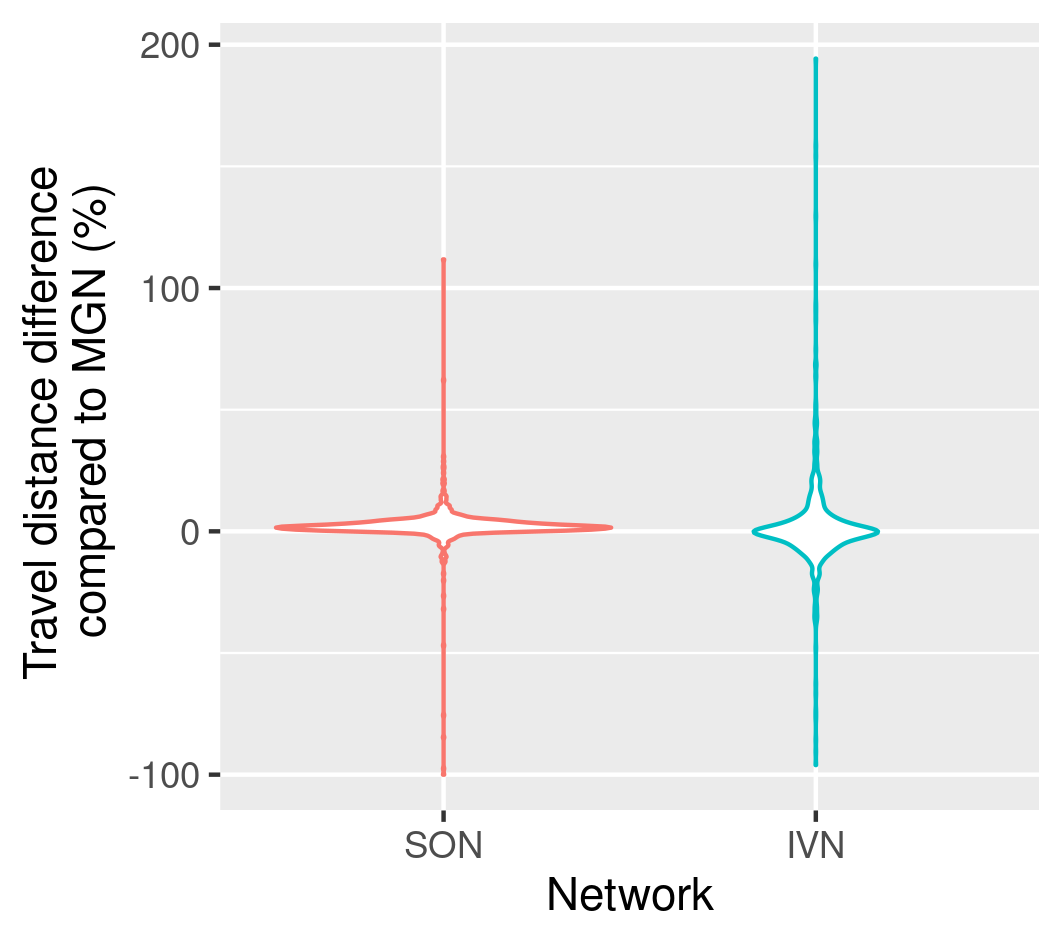}
        \caption{Cycling}
        \label{fig:distBike}
	 \end{subfigure}
   \begin{subfigure}[b]{0.45\textwidth}
        \includegraphics[width=\textwidth]{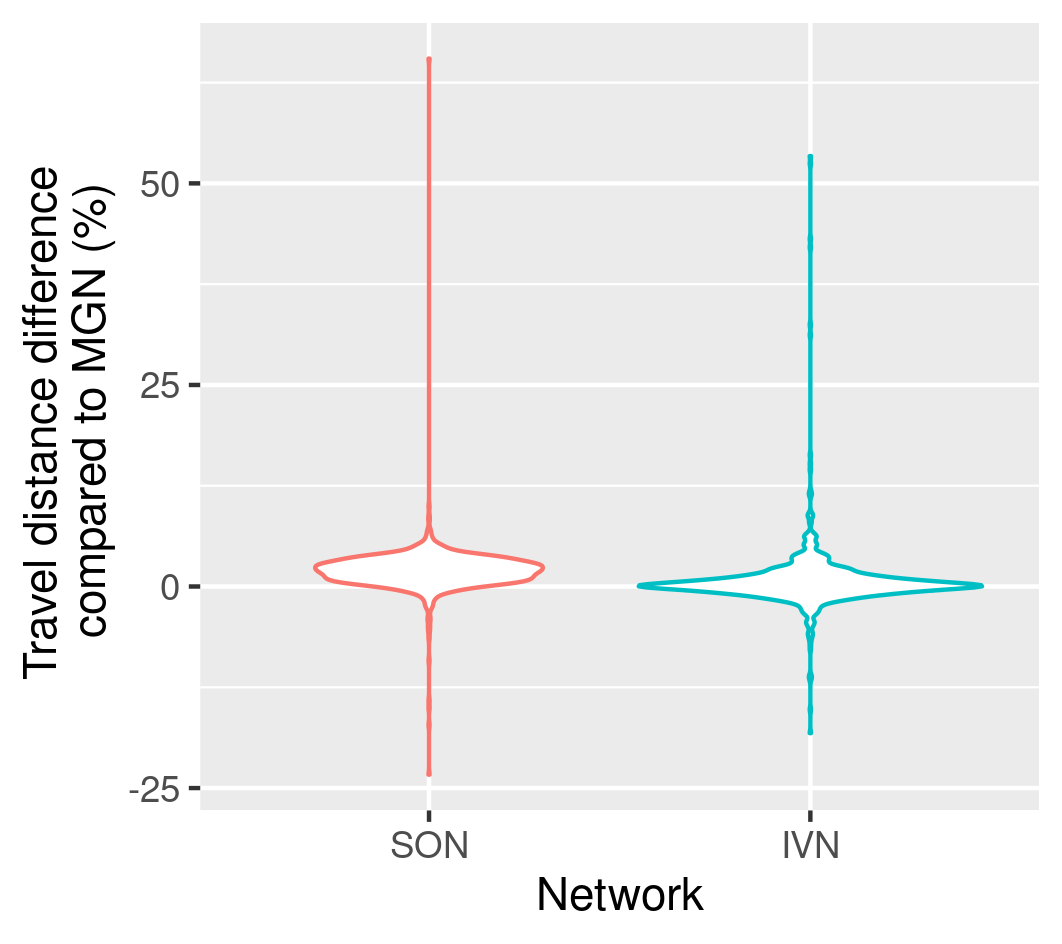}
        \caption{Driving}
        \label{fig:distCar}
	 \end{subfigure}
\caption{Travel distance comparison between different networks for walking, cycling and driving.}
\label{fig:comparisonPlots}
\end{figure}

Results of the shortest paths comparison of sample trips on \sn and \ivn relative to \mn (the baseline network) indicates that for walking, \sn is a considerably better option than using \ivn (Figure~\ref{fig:distWalk})\footnote{It should be noted that Infrastructure Victoria only used \ivn for car and public transport modelling.}. 
The average error percentage for walking, $E_{walk}$, when on \sn was $7.7\%$ $(SD=18.1)$ as compared with $15.6\%$ $(SD=28.8)$ for \ivn (Table~\ref{tab:pctDifferences}). 
Cycling trips also showed a considerably better fit to \mn when using \sn as compared with when using \ivn (Figures~\ref{fig:distBike}).
For \sn, as indicated in Table~\ref{tab:pctDifferences}, $E_{\sn}^{bicycle}$ was equal to $3.7\%$ $(SD=7.8)$ compared with $E_{\ivn}^{bicycle}=8.93\%$ $(SD=16.0)$ for \ivn. 

\begin{table}[ht]
\centering
\caption{Average trip distance from VISTA and average error percentages of travel distance of the shortest paths relative to \mn for trips on \sn,  $E_{\sn}$, and \ivn,  $E_{\ivn}$, for different travel modes.}
\label{tab:pctDifferences}
\begin{tabular}{@{}cccc@{}}
\toprule
            & VISTA travel distance, (SD)    & $E_{\sn}$, (SD)        & $E_{\ivn}$, (SD) \\ \midrule
    Walk    & 1.1km, (1.1)                   & 7.7\%, (18.1)          & 15.6\%, (28.8) \\
    Bicycle & 4.8km, (5.6)                   & 3.7\%, (7.8)           & 8.93\%, (16.0) \\
    Car     & 11.1km, (18.4)                 & 2.3\%, (2.6)           & 1.7\%, (3.8)   \\ 
 \bottomrule
\end{tabular}%
\end{table}

For car trips, however, the aggregated level comparison of Table~\ref{tab:pctDifferences} does not indicate a better fit in \sn than \ivn when compared with \mn.
This could be due to the road network design in Melbourne, which similar to most other Australian cities, has been traditionally car-oriented \citep{stevenson2016land}, meaning major straight roads connecting suburbs are common and therefore the shortest and fastest way to get from point A to B when trips are not local.  
Whereas for local walking and cycling trips that typically happen within a suburb, it was the local roads that provided the shortest paths.

\begin{figure}[ht]
    \centering
    \includegraphics[width=0.9\textwidth]{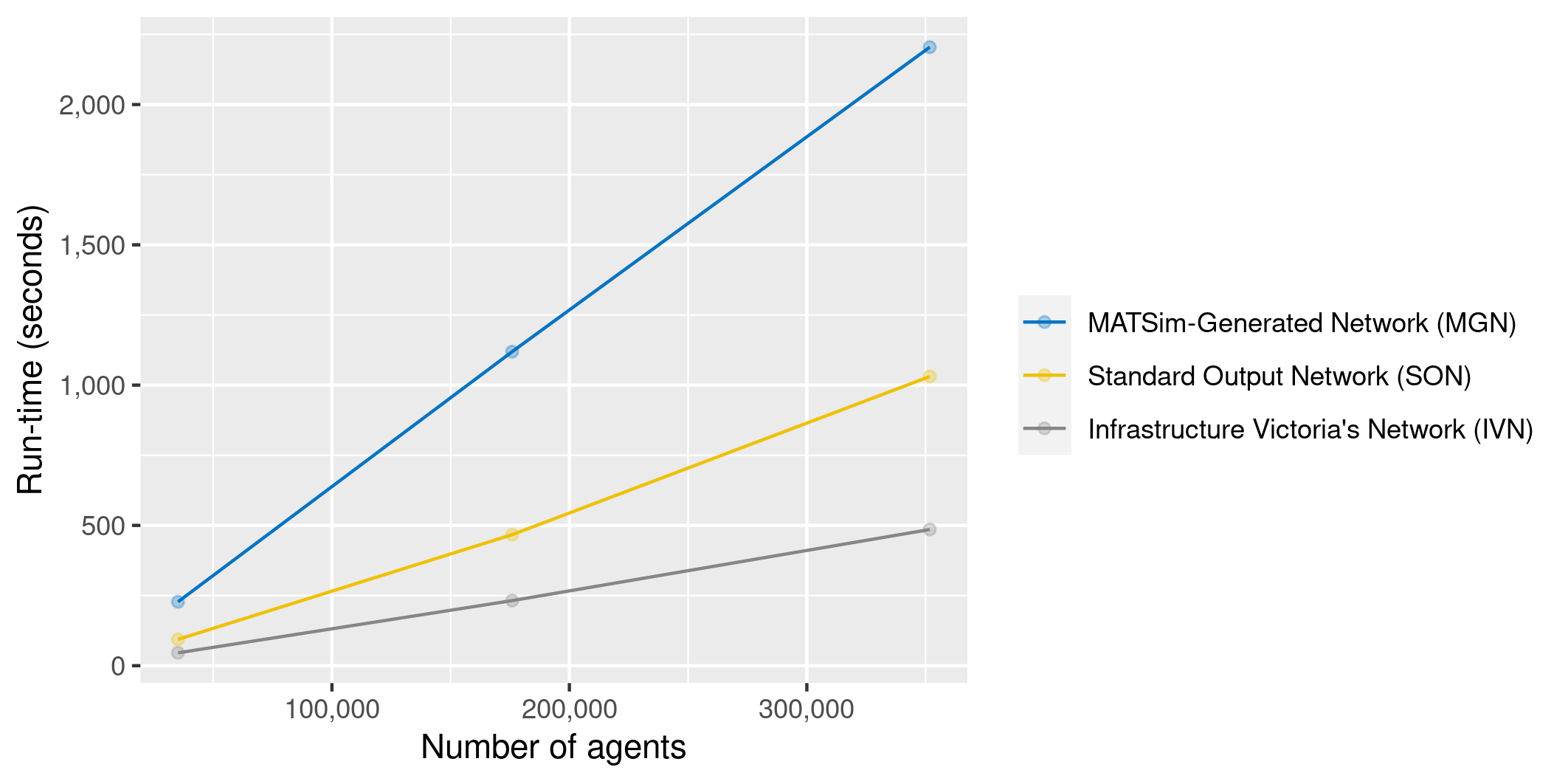}
    \caption{Average \ms iteration run-time for different networks and population sizes}
    \label{fig:simTime}
\end{figure}

Finally, the \ms average iteration run-time for different networks and sample sizes (Section~\ref{sec:simulation}) were compared with each other. 
As illustrated in Figure~\ref{fig:simTime}, iteration run-time was on average two times faster in \sn compared with \mn at all sample sizes.
This is due to the smaller number of links and nodes in \sn that results in less number of events (e.g., entering and exiting links) to be processed during the mobility simulation part of the \ms loop and also smaller search space and moving components to optimise for the re-planning step of \ms.
\ivn, given its considerably smaller size, was significantly faster than the other two networks, almost twice as fast as \sn and four times faster than \mn, which shows the usefulness of such a network when detailed roads are not critical.

\section{Discussion}
\label{sec:discussion}

In this paper, we presented our algorithm for building a road network suitable for active transportation modelling using open data.
\osm, as the main input of the algorithm, is a world-wide project with data available globally. \gls{gtfs} and \gls{dem}, as two additional inputs, are also publicly available for cities around the world.
Simulation-ready and well-maintained road network maps by governmental agencies, as a typical source for transport models (e.g., \citep{jacyna_modelling_2017,ryu_two-stage_2018}) are not frequent or easy to access for cities around the world, and even more uncommon are those incorporating active transport related attributes.
Furthermore, existing maps developed and maintained by different cities commonly have distinct formats and structure, which limits the applicability of a model developed for one city in other locations.
The work presented in this paper addresses this need by providing an automated and universal solution for road network generation for active transportation modelling.

The output network of the algorithm covers existing roads publicly accessible via walking, cycling or driving, with the optional step to add public transportation.
Despite the common practice of only including major roads in city-scale transportation models (e.g., \citep{infrastructure_victoria_model_2017}), walking and cycling necessarily involve a combination of major and minor roads when connecting individuals in residential locations to key destinations such as places of employment, education, social infrastructure or public transport.
By excluding minor roads, the resulting network also excludes efficient route options, since walkers and cyclists are to be routed along major roads only. 
Furthermore, the network with only major roads no longer represents how walkable an area truly is.
This matters given that a key environmental correlate of walking and cycling is high street connectivity, a measure consisting of the density of streets in a given area and a key component of a walkable area \citep{frank2005linking}. 
Street connectivity allows for greater route options potentially leading to more efficient travel times  \citep{sugiyama2012destination} as supported in the simulation modelling here through the more detailed \sn and \mn networks. 
The importance of low traffic neighbourhoods \citep{laverty2021low} and minor roads for walking \citep{gunn_designing_2017} and cycling \citep{broach_where_2012} is also supported in the literature, with minor roads particularly important for cyclists safety and riding preferences. 
Therefore, to have a proper road network for active transportation modelling, our algorithm captures all roads, major and minor, that pedestrians and cyclists are allowed to use.

Our algorithm incorporates key physical environment factors affecting active travel behaviour.
These factors include availability and type of the cycling infrastructure \citep{buehler_bikeway_2016}, road gradient \citep{hood_gps-based_2011} and intersections \citep{strauss_speed_2017}.
Safety and level of stress, as other key factors for cyclists' behaviour, although were not explicitly included, could be estimated using the information already captured by the algorithm following the approach similar to \citet{crist_fear_2019}.
There are, however, several other factors that are arguably also important for active travel behaviour that were not included, such as the width of bike paths and footpaths or on-street car parking \citep{chen_evaluating_2018}.
The decision to exclude such factors was due to data availability and our preference for using universally available data as the main inputs of the algorithm.
However, to address this limitation, the output network can be generated in popular GIS formats (i.e., ESRI Shapefile and SpatiaLite) that gives users the flexibility to perform further processing to add more layers of data from other sources.

Our network generation algorithm not only extracts the road geometries and their attributes but also simplifies the network to be suitable for using in city-scale simulation models where network size could be a key factor in determining the simulation run time \citep{waraich2015performance}.
Network simplification reduces the network size by approximately 77\% (from 1,660,750 nodes and links in \mn to 386,852 nodes and links in the \sn) as evident when comparing Figures~\ref{fig:osmInnMelb} and \ref{fig:allLinksInnMelb}.
As illustrated in Figures~\ref{fig:simTime}, this simplification results in a considerable reduction in simulation run-time overall -- approximately halving the run-time for the simulated sample.
This significant run-time saving was observed at various sample sizes. 

Simplifying the network could, however, impact the accuracy of the travel behaviours and the routes agents use, which required a more detailed examination.
Therefore, a comparison of the shortest paths for 1,000 randomly generated trips on each network and for each travel mode was undertaken.
The comparison of travel distance illustrates that the diversion from \mn, the baseline and most detailed network, was on average less than 10\% of all modes when using \sn (Table~\ref{tab:pctDifferences}). 
The minimum link length for the network simplification step (Section~\ref{sec:step3}) is an adjustable parameter (set to a default of 20m), giving users the flexibility to have more accurate representations of \osm depending on need.
However, it should be noted that decreasing the minimum link length increases the size of the network which results in longer simulation times.

A major roads only network, i.e., \ivn, used in the Victoria's state of art agent-based transportation model \citep{infrastructure_victoria_model_2017} was also examined in this paper.
As illustrated in Figure~\ref{fig:simTime}, \ivn provides a considerably faster simulation run-time compared with \sn and \mn -- approximately four times faster than \mn and two times faster than \sn. 
This considerable run-time speed reduction, however, comes at the cost of significant deviations from \mn in terms of travel distance for walking and cycling (Figure~\ref{fig:comparisonPlots}). 
Despite its shortcomings for simulating active transportation, for car traffic, however, no significant difference was observed between \ivn and \sn travel distance.

Therefore, if cars are the only travel mode of interest for a simulation model, using a simple network with only the major roads such as \ivn might be a reasonable option given its faster simulation times.
Networks such as \ivn could also be a reasonable options for when modelling public transportation given common \gls{pt} modes (i.e., train, tram and bus) either have their dedicated paths or travel on major roads.
However, as \gls{pt} trips usually require walking to and from the station, limitations of a simple network to capture this walking component should be taken into consideration.
Whereas if active transportation is important for the use case, we recommend using the \sn given it provides a middle point between run-time efficiency of \ivn and accuracy of \mn.
Our algorithm can be adjusted to create simple network major road only if desired by removing roads at \textit{Residential} level or lower from Table~\ref{tab_osmTagDefs} in the code\footnote{\url{https://github.com/matsim-melbourne/network/blob/master/functions/buildDefaultsDF.R}}. 

\begin{figure}[ht]
    \centering
    \includegraphics[width=0.6\textwidth]{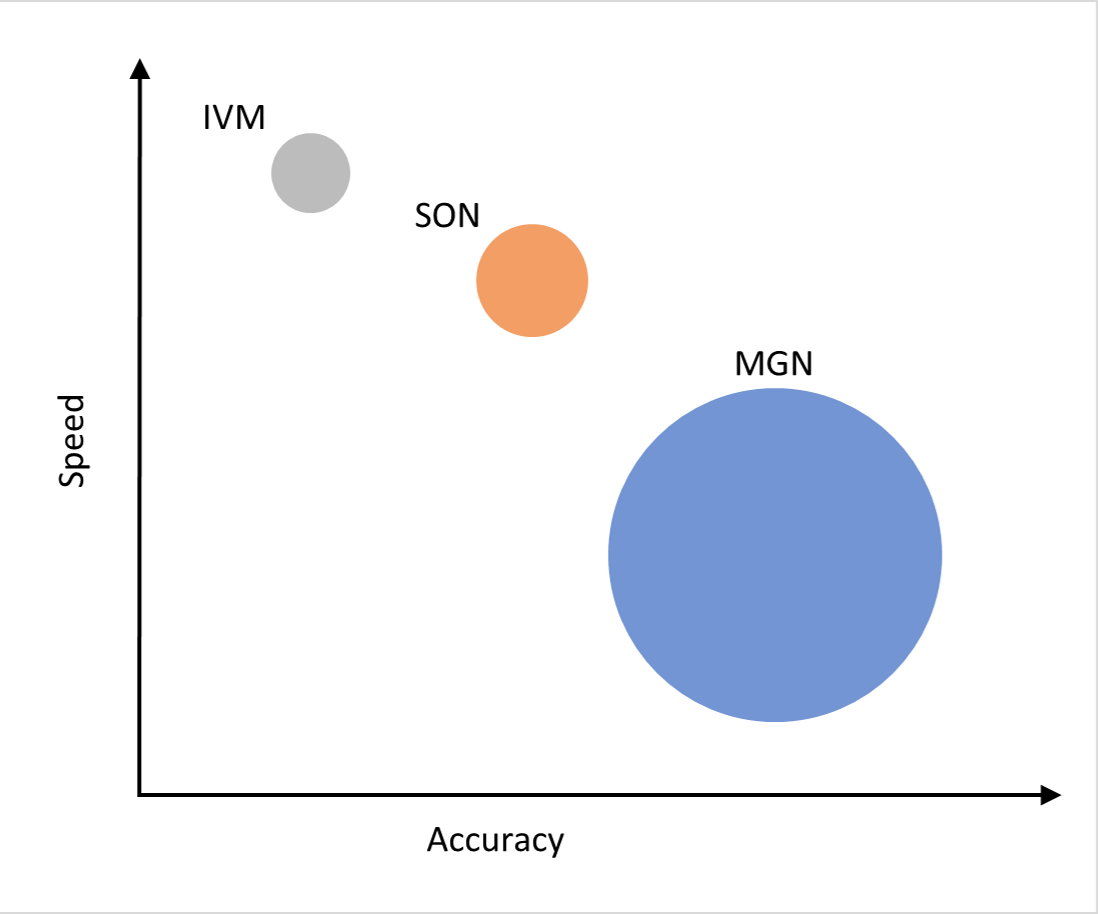}
    \caption{A schematic comparison of the our road network (\sn) to alternatives in terms of accuracy for active transport modelling (average $E$ for all modes), relative simulations run-time speed, and relative network size.}
    \label{fig:compare}
\end{figure}

Figure~\ref{fig:compare}, provides a schematic comparison between the three networks we covered in this paper, (\sn, \mn and \ivn) in term of the average overall accuracy for all modes, run-time speed and the network size. 
As illustrated in Figure~\ref{fig:compare}, our network, \sn, provides a balanced alternative to the two common practices of major roads only networks, \ivn, and exact \osm conversion networks, \mn.

\begin{figure}
    \centering
    \includegraphics[width=0.7\textwidth]{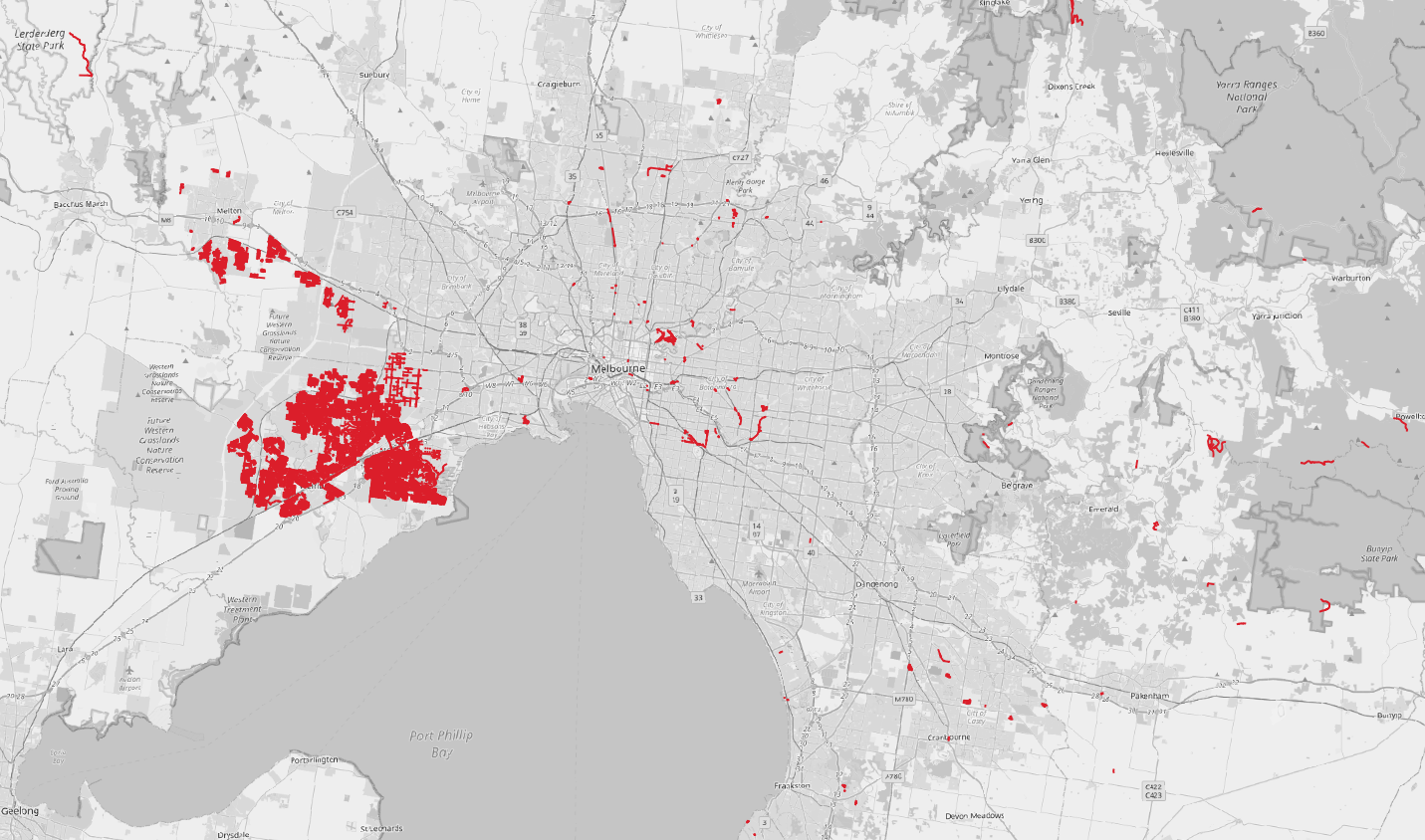}
    \caption{Distribution of \osm links with the \texttt{smoothness} attribute value available, Greater Melbourne (base map from \textcopyright\href{https://www.openstreetmap.org/copyright}{OpenStreetMap}).}
    \label{fig:smoothness}
\end{figure}

We used \osm as the main input of our algorithm.
Although this provides flexibility and universality for the algorithm, relying on \osm comes with a number of limitations and considerations.
A key limitation of relying on \osm is that tagging approaches might differ between regions. 
For example, \citet{ferster_using_2020} found inconsistencies in the use of \texttt{highway=cycleway} tag in Canada as it was referring to both off-road bike paths and on-street separated bike lanes.
Furthermore, some tags might be more common in some areas and not others. 
\citet{ziemke_bicycle_2018} found that only 12\% of the \osm links in Berlin had the tag \texttt{smoothness} that can be used to evaluate cycling comfort. For Melbourne, as shown in Figure~\ref{fig:smoothness}, the \texttt{smoothness} tag was mostly used in the south-western suburbs, which indicates that inconsistencies can exist even within one city.

The focus of the algorithm was to provide a flexible tool for building road networks for active transport models. 
Therefore, only a selected number of more commonly used road attributes was covered by the algorithm.
There are some other useful and relevant tags and information in \osm that could be added to the algorithm as next steps such as bicycle parking, \texttt{amenity=bicycle\_parking}, or pedestrian crossings, \texttt{crossing=marked AND highway=crossing}.
These were not included in this version of the algorithm as there are still high inconsistencies in tagging them, however, it is expected the quality of these tags to improve gradually as more people are becoming interested in walking and cycling data and to become suitable for including in algorithms such as ours in near future.

 \section{Conclusion}
 
In this paper, an algorithm for building the road network description to be used in city-scale active transportation simulation modelling was introduced.
The algorithm covers common road types publicly accessible via car, walking and cycling. 
It also extracts the key road attributes required for modelling active transportation and also simplifies the network to minimise the impact of a detailed network on simulation run-time.

A comparison between the non-simplified network, \mn, our algorithm's standard output (minimum link length=20m), \sn, and a major roads only road network, \ivn, showed that our algorithm offers a good balance between accuracy and performance for walking and cycling modes.
Our proposed algorithm provides an alternate solution for traditional city-scale models that typically cover only major roads, towards detailed networks suitable for modelling active transportation.

\section*{Acknowledgement}
This paper is part of the development of a city-scale \ms model for Melbourne. 
We thank the MATSim Melbourne community and our colleagues at Healthy Liveable Cities Group, Centre for Urban Research, RMIT University for their inputs and comments that informed this research. 
We would also like to show our gratitude to Infrastructure Victoria for sharing their agent-based transportation model with us, we used its network for comparison in this paper.

\section*{Funding declaration}
AJ is supported by an Australian Government Research Training Program Scholarship.
DS's time on this project is funded by Collaborative Research Project grants from CSIRO's Data61 (2018-19, 2020-21).
LG's time on this project is funded by the NHMRC funded Australian Prevention Partnership Centre (\#9100001).

\bibliographystyle{apalike}
\bibliography{references.bib}

\end{document}